\documentclass{aa}  

\usepackage{natbib}
\usepackage{graphicx}
\usepackage{subfigure}
\usepackage[varg]{txfonts}
\usepackage[colorlinks=true,linkcolor=blue,citecolor=blue,filecolor=blue,urlcolor=blue]{hyperref}  
\usepackage{CJKutf8}

\usepackage{hyperref}

\usepackage{graphicx}	
\usepackage{amsmath}	
\usepackage{amssymb}	
\usepackage[version=4]{mhchem}

\makeatletter
\def\uwave{\bgroup \markoverwith{\lower3.5\p@\hbox{\sixly \textcolor{red}{\char58}}}\ULon}
\font\sixly=lasy6 
\makeatother

\usepackage{color}
\definecolor{mygray}{gray}{0.6}
\definecolor{orange}{rgb}{1.0, 0.4, 0.0}
\usepackage{bm}
\usepackage{color}
\usepackage{ulem}
\usepackage{xspace}
\newcommand{\Se}[1]{Section~\ref{sec:#1}}

\newcommand{\Fg}[1]{Figure~\ref{fig:#1}}
\newcommand{\fg}[1]{Fig.~\ref{fig:#1}}

\newcommand{\Tb}[1]{Table~\ref{tab:#1}}
\newcommand{\eq}[1]{Eq.~(\ref{eq:#1})}
\newcommand{\Eq}[1]{Equation~(\ref{eq:#1})}
\newcommand{\Eqs}[2]{Equations~(\ref{eq:#1}) and (\ref{eq:#2})}
\newcommand{\codename}{\texttt{ExoLyn}\xspace}

\usepackage{ulem}
\makeatletter
\def\uwave{\bgroup \markoverwith{\lower3.5\p@\hbox{\sixly \textcolor{red}{\char58}}}\ULon}
\font\sixly=lasy6 
\makeatother

\begin{document}

    \title{\codename: a golden mean approach to multi-species cloud modelling in atmospheric retrieval} 
    \author{Helong Huang (\begin{CJK*}{UTF8}{gbsn}黄赫龙\end{CJK*})
           \inst{1}
           \and
           Chris W. Ormel\inst{1}
           \and
           Michiel Min\inst{2}
           }
    \institute{
        Department of Astronomy, Tsinghua University, Haidian DS 100084, Beijing, China \\
            \email{huanghl22@mails.tsinghua.edu.cn}
            \and
        SRON Netherlands Institute for Space Research, Niels Bohrweg 4, 2333CA Leiden, The Netherlands
        }

\authorrunning{Huang, Ormel, Min}
\date{Accepted XXX. Received YYY; in original form ZZZ}

\abstract{Clouds are ubiquitous in exoplanets' atmospheres and play an important role in setting the opacity and chemical inventory of the atmosphere. Understanding clouds is a critical step in interpreting exoplanets' spectroscopic data.}
{The aim is to model the multi-species nature of clouds in atmospheric retrieval studies. To this end, we develop \codename\ -- a 1D cloud model that balances physical consistency with computational efficiency.}{\codename solves the transport equation of cloud particles and vapor under cloud condensation rates that are self-consistently calculated from thermodynamics. \codename is a standalone, open source package capable to be combined with \texttt{optool} to calculate solid opacities and with \texttt{petitRADTRANS} to generate transmission or emission spectra.}
{With \codename we find that the compositional structure of clouds in hot Jupiter planets' atmospheres is layered with a cloud dominated by magnesium-silicates on top of an iron cloud. This finding is consistent with more complex cloud formation models but can be obtained with \codename in only a few seconds. The composition of the cloud particles can be constrained from the spectrum, for example, \ce{MgSiO3} and \ce{Mg2SiO4} components give rise to an absorption feature at $8-10\ \mathrm{\mu m}$. We investigate the dependence of the cloud structure on the bulk elemental composition of the planet and find that \ce{SiO2}-dominated clouds forms on metal-rich planet and \ce{Fe} clouds with strong extinction effect forms on C-rich planet.} 
{Designed towards maximum flexibility, \codename can also be used in retrieval analysis of sub-Neptunes and self-luminous planets. The efficiency of \codename opens the possibility of joint retrieval of exoplanets' gas and cloud components. }

\maketitle


\section{Introduction}

Transmission and emission spectra are powerful diagnostics to obtain the composition of a planet atmosphere. From the wavelength and strength of the spectral features, the presence of a variety of molecules has been identified and their abundance has been constrained. 
With HST and Spitzer, one of the most well-observed molecules in exoplanet spectrum is \ce{H2O} \citep[e.g.,][]{HuitsonEtal2013, DemingEtal2013}. Before JWST, the detection of carbon-bearing molecules, like \ce{CO} and \ce{CH4}, are mostly achieved by emission spectroscopy and are limited to giant planets \citep{KonopackyEtal2013, BarmanEtal2015, SheppardEtal2017}. However, the constraint on the abundance of C- or O-bearing species aside from \ce{H2O} is limited by wavelength coverage and sensitivity \citep{Madhusudhan2019}. 
With its unprecedented spectral resolution, James Webb Space Telescope (JWST) has opened a new era for atmospheric spectrum characterization. Covering the spectral range from $0.6\ \mathrm{\mu m}$ to $15\ \mathrm{\mu m}$, JWST has robustly observed \ce{CO} \citep[e.g.][]{RustamkulovEtal2023}, \ce{CO2} \citep[e.g.][]{GrantEtal2023, XueEtal2024, AldersonEtal2023} and \ce{CH4} \citep[e.g.][]{BellEtal2023} on giant planets using transmission spectroscopy. 
Moreover, JWST has extended our spectroscopic understanding to super-Earth planets. It detected signals of \ce{CO2} and \ce{CH4} on hycean candidates TOI-270 d \citep{HolmbergMadhusudhan2024} and K2-18 b \citep{MadhusudhanEtal2023}.
For terrestrial planets, the emission spectrum of 55 Cnc e revealed a feature from \ce{CO} or \ce{CO2} \citep{HuEtal2024}. Besides gas features, solid phase silicate features have been discovered on hot Jupiters like WASP-107 b \citep{DyrekEtal2024} and WASP-17 b \citep{GrantEtal2023} for the first time. 

The composition of an exoplanet is the key indicator of its formation location and evolution pathway. This is because the composition of the building blocks the planet has accreted varies with the local physical conditions (temperature, pressure) in the protoplanetary disk. For example, the location of the snowline regulates the existence of volatiles and the content of the corresponding vapor \citep{ObergEtal2011}. 
The ratio and composition of solid and gas material which a planet accretes, determines the bulk properties of the planet, like metallicity, C-to-O ratio (C/O) and volatile ratio \citep{MadhusudhanEtal2014, BoothEtal2017, SchneiderBitsch2021, TurriniEtal2021, KhorshidEtal2022, DantiEtal2023}. 
In particular, the metallicity and C/O are identified as important tracers of planet formation history, because C- or O- bearing molecules are most commonly constrained in atmospheres. 
In addition, N/O or S/N offer valuable indicators to constrain a planet's formation history \citep{TurriniEtal2021, KhorshidEtal2024}. JWST has detected \ce{SO2} on a giant exoplanet atmosphere \citep{AldersonEtal2023, DyrekEtal2024}, providing the opportunity to trace back their formation \citep{SchneiderBitsch2021, Crossfield2023, KhorshidEtal2024}. This is because the main sulfur carrier, \ce{FeS}, condenses at a temperature higher than carbon or oxygen ($650\ \mathrm{K}$) \citep{KamaEtal2019}, which indicates that S has been accreted in solid form. 

Inferring the chemical makeup and the abundance of an atmosphere from spectroscopic data is a process referred to as ``retrieval''. In retrieval a parameter-dependent forward model is computed to obtain a synthetic spectrum that is compared to the observational data.
When performing retrieval, Bayesian analysis like Markov-chain Monte Carlo \citep[MCMC, e.g.][]{MadhusudhanEtal2011, BennekeSeager2012} or nested sampling method \citep[e.g.][]{LineParmentier2016, LavieEtal2017} are applied on the observed spectrum, to test how different sets of parameters fit the data. Bayesian analysis is done by running the forward model thousands to millions times \citep[e.g.][]{ChubbMin2022} to obtain the probability distribution of each parameter.  As the parameter space is highly multi-dimensional and the forward computation is repeated many times, the CPU runtime of each forward run must be short, preferentially limited to the order of seconds.
There are already a dozen of retrieval code to date \citep{MacDonaldBatalha2023}, for example \texttt{NEMESIS} \citep{IrwinEtal2008}, \texttt{TauREx} \citep{Al-RefaieEtal2021}, \texttt{POSEIDON} \citep{MacDonaldMadhusudhan2017}, \texttt{ATMO} \citep{WakefordEtal2017}, \texttt{petitRADTRANS} \citep{MolliereEtal2019}, \texttt{ARCiS} \citep{MinEtal2020}, \texttt{PICASO} \citep{MukherjeeEtal2021}, with varying assumptions and physical complexity. 
This complexity occurs in many aspects of the forward model, like gas-phase chemistry, temperature profile, or molecular opacity model. For example, free-chemistry, where molecular abundances are free parameters, and equilibrium chemistry, in which the Gibbs free energy is at its minimum, are two common choices for gas-phase chemistry.

A particular aspect atmosphere forward models need to address is the treatment of clouds. Clouds are ubiquitously found on exoplanets and brown drawfs \citep[e.g.][]{PontEtal2008, PontEtal2013, KreidbergEtal2014}. Clouds affect the spectrum in a variety of ways. First, because they obscure the deep atmosphere from view, they tend to smooth out molecular features. 
At near-IR-to-visible wavelengths, small cloud particles or hazes high up in the atmosphere can lead to scattering slopes \citep{PontEtal2013, SedaghatiEtal2017}. Second, at mid-IR, clouds display distinct solid features (e.g., the $10\ \mathrm{\mu m}$ silicate feature; \citealt{MilesEtal2023, DyrekEtal2024}). Third, the formation of clouds consumes condensable vapor, which thereby depletes the corresponding atoms in the atmosphere \citep{KomacekEtal2022, LeeEtal2023}. The change in the molecular inventory is reflected in the spectra by weaker molecular features. Moreover, the presence of clouds can also change the heat transfer, influencing the temperature structure in the atmosphere \citep{RomanRauscher2019, HaradaEtal2021, RomanEtal2021}. Therefore, unless atmospheres are undeniably cloud-free, understanding clouds is essential to interpret exoplanet spectral data.

Most forward models in retrieval codes parameterize clouds (see \citealt{Barstow2020} for an overview of cloud parameterizations) In its simplest form, this entails a cloud deck pressure below which the atmosphere is optically thick \citep{BennekeSeager2012}. Some models design a parameterized function representing cloud opacity, while others impose monodisperse particles and calculate Mie scattering \citep{LavieEtal2017}. Importantly, in these models the properties of the cloud layers are not limited by physical principles. 
A more physically motivated cloud model is proposed by \citet{AckermanMarley2001} (AM01 hereafter). This model assumes that all saturated material condense to solid, and solves for the condensate concentration by balancing the upward diffusive motion with the raindrop sedimentation.  The AM01 cloud model is used extensively in retrieval codes, e.g. \texttt{petitRADTRANS} and \texttt{PICASO}. 
While popular, the AM01 model relies on ad-hoc parameterizations. The particle sedimentation velocity is prescribed with an $f_\mathrm{sed}$ parameter and the particle sizes are not solved for. 

Other cloud models are physically motivated.
\texttt{DRIFT} \citep{HellingWoitke2006, HellingEtal2008} solves the moment equations of the particle size distribution. Nucleation, condensation, coagulation and settling are modeled in detail. Further complication comes from computing the time-dependent particle size distribution. This approach is adopted by \texttt{CARMA} \citep{TurcoEtal1979, ToonEtal1979}, and is applied to \ce{H2SO4} clouds on Venus and silicate clouds on hot Jupiters \citep{PowellEtal2018}.
Recently \citet{KieferEtal2024} studied time-dependent cloud formation under gas phase disequilibrium, and found nucleation and cloud formation can happen within 1 second. 
Though these models consider a variety of physical processes, and promote our understanding on exoplanet clouds, they consume significantly more computational resources than parameterized cloud models. 
Therefore, forward model with high level of physical complexity is usually ran on a limited number of parameter grids \citep[e.g.][]{MukherjeeEtal2021}, rather than in the MCMC or nested sampling simulations.

The \texttt{ARCiS} atmosphere model \citep{MinEtal2020, OrmelMin2019} attempts to balance physical consistency with computational efficiency. The philosophy of \texttt{ARCiS} is to model physical processes that affect the cloud formation most, while parameterizing processes for which we have a poor understanding or those that are too time-consuming for a robust retrieval.
This approach has been successfully applied to retrievals using self-consistent cloud formation \citep{MinEtal2020}, disequilibrium chemistry \citep{KawashimaEtal2019}, and 3D phase curve retrievals \citep{ChubbMin2022}. However, one aspect where \texttt{ARCiS} (and other) cloud models can be improved on is the multi-species nature of clouds. 

It is indeed expected that the composition of the cloud particles changes as function of atmospheric depth along with the changing molecular inventory and physical conditions (temperature, pressure). Therefore, an efficient cloud model that simulates, to first order, the realistic condensation and transport process taking place in the atmosphere would be a great asset to the community. To this effect, we have developed a multi-species cloud model, \codename, which computes cloud condensation, sedimentation and diffusion in a computationally efficient, yet physically consistent way. 
The input for \codename is a temperature-pressure (T-P) and nucleation profile, cloud forming reactions and  vapor abundances at the lower atmosphere boundary. Then \codename will compute the cloud concentration and components by solving the transport equation including condensation, sedimentation and diffusion. 
\codename has been designed to study cloud formation across a variety of planets: hot Jupiters, sub-Neptunes, and self-luminous planets. \codename is publicly available\footnote{\url{https://github.com/helonghuangastro/exolyn}} , written mainly in \texttt{Python}, and compatible with popular retrieval frameworks like \texttt{petitRADTRANS}. 

This paper is structured as follows. In \Se{method} we review the \texttt{ARCiS} cloud model and extend it to multiple condensate species. In \Se{result} we present the \codename results for hot Jupiter planets. Our results are benchmarked against previous models in \Se{comparison}. The resulting transmission spectra are shown in \Se{spectrum}. The application of \codename to self-luminous planets and sub-Neptunes are demonstrated in \Se{application} to show the code's flexibility. 
In \Se{discussion}, \codename's performance is assessed and its implication to future atmospheric retrieval is discussed. Finally we summarize the paper in \Se{conclusion}.

\section{Method} \label{sec:method}
\codename extends the single species cloud model by \citet{OrmelMin2019} (OM19 hereafter). In \codename, a cloud particle is composed of one or more nuclei and multiple condensate species. The condensates form following actual chemical reactions, which we will describe in detail in \Se{method-multi}. In \Se{singlemethod} we first introduce clouds made of nuclei and a single condensate species.
\subsection{Single condensate species cloud} \label{sec:singlemethod}
We briefly introduce the OM19 1D cloud model  which is dedicated to a single condensate cloud species.
OM19 considered a cloud particle composed of a nucleus, coated by single-species condensate formed from corresponding hypothetical vapor (e.g, \ce{MgSiO3} vapor). 
The condensate is formed on the nuclei according to the saturation ratio, $S=p_\mathrm{vap}/p_\mathrm{sat}$, where $p_\mathrm{vap}$ is the vapor partial pressure 
and $p_\mathrm{sat}$ is its saturation pressure. Besides condensation, particles also grow by colliding and sticking with other particles. 
After formation, cloud particles are transported through the cloud by settling and turbulence, characterized by a turbulent diffusion coefficient $K_{zz}$. 
The combination of condensation, diffusion, settling and coagulation is solved for a steady solution, where all processes are in equilibrium.
Let $x_c$ and $x_v$ to be the mass mixing ratio of the condensate and vapor species, the transport equations for condensate and vapor are
\begin{eqnarray}
    -\frac{\partial}{\partial z}\left(K_{zz}\rho_\mathrm{gas}\frac{\partial x_c}{\partial z}\right) + \frac{\partial}{\partial z}\left(x_c\rho_\mathrm{gas}v_\mathrm{sed,p}\right) = \mathcal{S}_{c} \label{eq:condensatesingle}\\
    -\frac{\partial}{\partial z}\left(K_{zz}\rho_\mathrm{gas}\frac{\partial x_v}{\partial z}\right) = -\mathcal{S}_c. \label{eq:vaporsingle}
\end{eqnarray}
In the equations above, $\rho_\mathrm{gas}$ is the total gas density. The left hand side contains the transport terms and the right hand side of \eq{vaporsingle} is the source or sink term for condensates and vapor, respectively.
The particle sedimentation (gravitational settling) velocity $v_\mathrm{sed, p}$ is assumed to be the terminal velocity according to the Epstein drag law and the Stokes drag law \citep{Weidenschilling1977}, i.e. $v_\mathrm{sed, p} = a_p \rho_\bullet g \max \{1, 4a_p/9\lambda\}/ v_\mathrm{th} \rho_\mathrm{gas}  $, 
where $a_p$ is the particle radius, $\rho_\bullet$ is the particle density, $g$ is gravitational acceleration, $\lambda$ is the gas mean free path and $v_\mathrm{th}$ is gas thermal velocity.

At each location, we assume the particles have a single size. In addition to the solid mixing ratio, the particle number density is required to calculate the particle size, which determines the sedimentation velocity and the condensation rate $\mathcal{S}_c$.
Therefore another equation for the particle number density $n_p$ is needed to supplement the set of \Eqs{condensatesingle}{vaporsingle}. To keep  all of the unknown variables dimensionless, OM19 solved for the ``nuclei mixing ratio'' defined as particle number density times nucleus mass normalized by gas density, $x_n = n_p m_{p0} / \rho_\mathrm{gas}$ , where $m_{p0}$ is the mass of the nuclei particles.
Besides diffusion and settling, collision among the nuclei results in a sink term for their number density:
\begin{equation}
    -\frac{\partial}{\partial z}\left(K_{zz}\rho_\mathrm{gas}\frac{\partial x_n}{\partial z}\right) + \frac{\partial}{\partial z}\left(x_n\rho_\mathrm{gas}v_\mathrm{sed,p}\right) = \left(\mathcal{S}_n-\frac{x_n\rho_\mathrm{gas}}{t_\mathrm{coag}}\right), \label{eq:nuclei}
\end{equation}
where $\mathcal{S}_n$ is the nucleation rate and $t_\mathrm{coag}$ is the coagulation time scale. 
The nucleation rate profile depends on the nuclei species \citep{GailEtal1984, LeeEtal2015}, curvature of nuclei (the so-called Kelvin effect) and temperature profile \citep{LeeEtal2018} and is highly uncertain (see review by \citet{Helling2019}). Thus, instead of deriving $\mathcal{S}_n$ from first principles, OM19 took a prescribed log-normal distribution along the z direction, 
where the location of nucleation maxima, the spread of log-normal distribution and the total amount of nucleation are prescribed as parameters (see \Tb{parameters}).
With solid and nuclei mass mixing ratio in hand, the mass for a single particle is $m_p = m_{p0} \times x_c / x_n$. The particle radius is found from its mass and internal density $\rho_\bullet$, assuming a spherical shape. Thereafter, the particle coagulation time scale $t_\mathrm{coag}$ can be calculated with Equation (12) of OM19:
\begin{equation}
    t_\mathrm{coag}^{-1} = \frac{1}{2}\pi n_p (2a_p)^2 \Delta v + 2\pi n_pa_p\min{\left(v_\mathrm{BM}a_p, D_p\right)},
\end{equation}
where $v_\mathrm{BM} = \sqrt{16k_\mathrm{B}T/\pi m_p}$ is the brownian motion velocity, $D_p=k_\mathrm{B}T/6\pi\nu_\mathrm{mol}\rho_\mathrm{gas}a_p$ and $\Delta v = v_\mathrm{sed, p}/2$ comes from differential settling \citep{OkuzumiEtal2012}. $\nu_\mathrm{mol}$ is the molecular viscosity.

For boundary condition, OM19 fixed the solid and nuclei concentration at the lower boundary to be zero, and the vapor concentration to be the envelope bulk concentration, as every cloud species should sublimate for hot planet with substantial envelopes at sufficient depths. 
For the upper boundary, OM19 fixed the solid, vapor and nuclei fluxes across it to be 0, as the planet is assumed to be isolated and static, without material flowing into or escape from it. 

The single species cloud model in OM19 is generalized to multiple species by \citet{MinEtal2020}. This is done by running the cloud formation model once at a time for each cloud species, and iterating over all cloud species. 
However the problem arises that the sequence of iteration may matter in the final cloud concentration, because the species condensing first has access to all gas phase molecules, leaving less reactant for the next condensate. 
In this work we overcome this problem by calculating the condensation rate for all condensate species in a thermal chemical approach, which is described in \Se{method-multi}. 

\subsection{Cloud with multiple condensates}
\label{sec:method-multi}
\begin{figure}
    \includegraphics[width=\columnwidth]{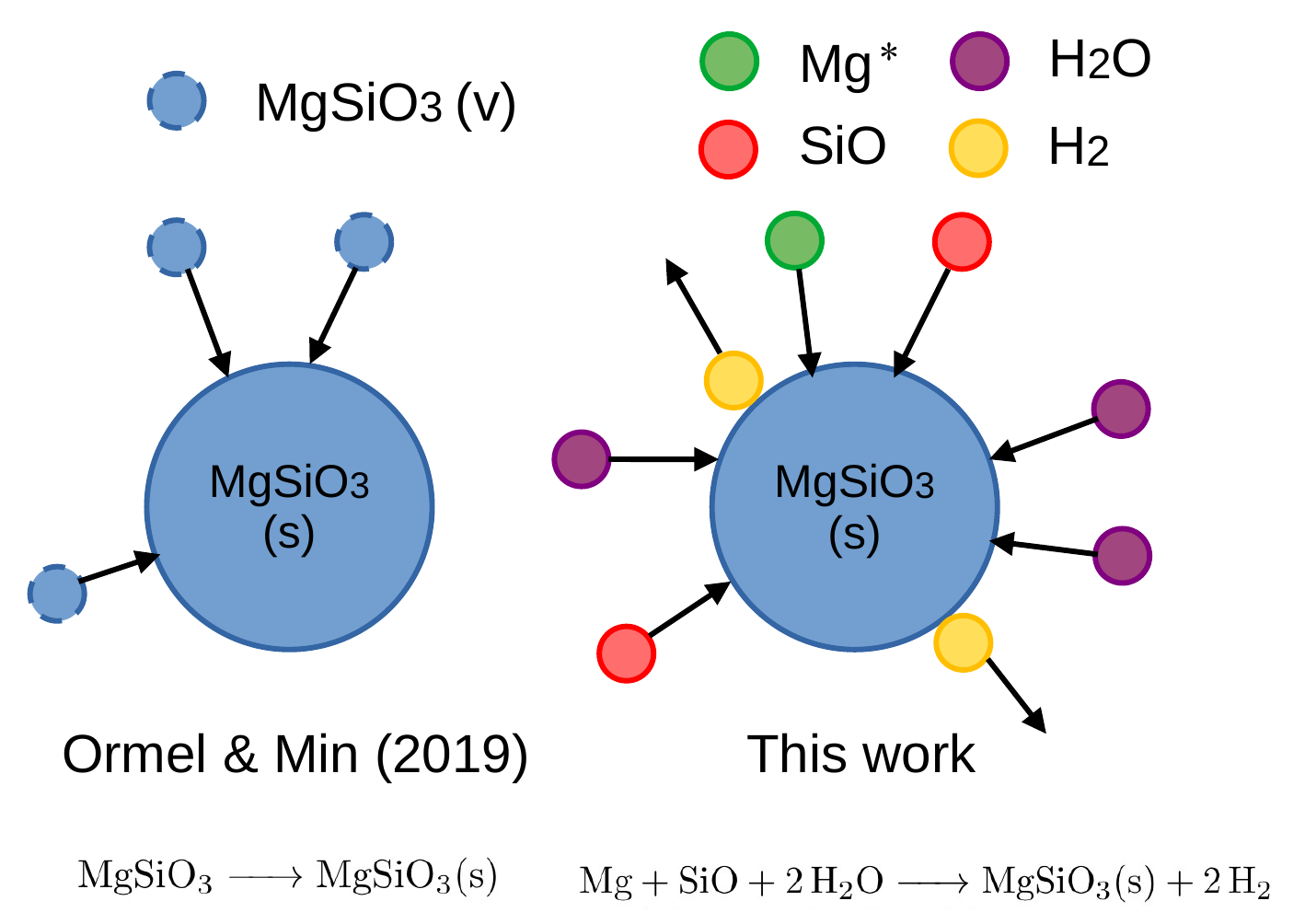}
    \caption{\label{fig:cartoon} 
    Comparison of \codename's multi-species approach to condensation against the single-species approach of \citet{OrmelMin2019}. \codename considers the actual vapor species that are present in the atmosphere. As illustrated in the right panel, the condensation rate of \ce{MgSiO3} is determined by the amount of \ce{Mg} -- the species that limits the condensation by its vapor concentration and stoichiometric number in the reaction.}
\end{figure}

We extend the cloud formation model to multiple cloud components. 
Instead of assuming the presence of hypothetical vapor in the gas phase (e.g. \ce{MgSiO3} vapor), we consider the realistic chemical reaction from which multiple vapor species condense. 
The difference between the condensation reaction in our work and OM19 is shown in \Fg{cartoon}, where we take the formation reaction of \ce{MgSiO3} as an example, 
\begin{equation}
\ce{Mg + SiO + 2H2O -> MgSiO3(s) + 2H2}.    \label{eq:MgSiO3}
\end{equation}
Let subscript $i$, $j$ and $k$ denote the $i^\mathrm{th}$ solid phase species, the $j^\mathrm{th}$ vapor species and the $k^\mathrm{th}$ chemical reaction, the general form of which is 
\begin{equation}
    \label{eq:reaction}
    \sum_{\mathrm{all\ vapor\ }j\mathrm{\ involved\ in\ }k} \nu_{k, j} V_{j} \rightarrow  S_{i(k)},
\end{equation}
where $\nu_{k, j}$ is the stoichiometric number of vapor species $V_j$ in the $k$-th reaction. In \Eq{reaction} we have normalized the stoichiometric numbers with respect to that of the solid. Thus $\nu_{k, j}$ can be fractional. 
With this notation the stoichiometric number of any vapor product is negative, for example $\nu_{k, \ce{H2}}=-2$ for \Eq{MgSiO3}. The transport equations for condensate and vapor can be generalized as
\begin{eqnarray}
    -\frac{\partial}{\partial z}\left(K_{zz}\rho_\mathrm{gas}\frac{\partial x_{c, i}}{\partial z}\right) + \frac{\partial}{\partial z}\left(x_{c, i}\rho_\mathrm{gas}v_\mathrm{sed,p}\right) = \sum_k \alpha_{k, i} \mathcal{S}_{c, i(k)} \label{eq:condensate}\\
    -\frac{\partial}{\partial z}\left(K_{zz}\rho_\mathrm{gas}\frac{\partial x_{v, j}}{\partial z}\right) = -\sum_{k} \frac{\nu_{k, j}\mu_{j}}{\mu_{i(k)}} \mathcal{S}_{c, i(k)}, \label{eq:vapor}
\end{eqnarray}
where $\mu$ is the molecular weight of vapor ($\mu_j$) or condensate ($\mu_i$), $i(k)$ is the condensate formed in reaction $k$ and $\mathcal{S}_{c, i(k)}$ is the condensate i formation rate of reaction $k$. When reaction $k$ does not contribute to condensate $i$, $\alpha_{k, i}$ is 0, otherwise $\alpha_{k, i}$ is 1. 
 The nuclei mixing ratio $x_n$, proportional to the cloud particle number by definition, is still described by \Eq{nuclei} and the mass of a single particle is updated according to $m_p = m_{p0} \times \Sigma_i x_{c, i} / x_n$. 
Note that we do not follow the concentration of gas which is abundant in the atmosphere, e.g. \ce{H2}, \ce{He}, or which does not participate in cloud formation, like \ce{CO}. Their combined concentration is defined as $x=1$, with mean molecular weight $m_\mathrm{gas}$. With this assumption, the production or consumption of vapor does not change $m_\mathrm{gas}$.

Particles grow because vapor condenses on their surfaces. For a given condensate species, there exists one vapor molecule that limits the growth of the condensate (see \Fg{cartoon}), which we refer to as the ``limiting vapor species'', denoted as $j^\ast$ for reaction $k$.
Take, for example, the formation of \ce{MgSiO3} (\Fg{cartoon}). If the impinging rate of \ce{Mg} is 1 per unit time, \ce{SiO} is 2 and \ce{H2O} is 3, then the reactant limiting the reaction rate is \ce{Mg}.
Formally, the limiting species is determined by finding the vapor species $j$ that minimizes $\{x_{v, j}v_j/\mu_j\nu_{k,j}\}$, for all vapor species taking place in reaction $k$. 
We follow \citet{WoitkeHelling2003} for the impinging velocity $v_{j^\ast}=\min\{v_{\mathrm{th}, j^\ast}, 4 D/a_p \}$, where $D$ refers to the molecular diffusivity.
Generalizing Equation (9) in OM19, the condensation rate $\mathcal{S}_{c, k}$ can be expressed as
\begin{equation}
    \label{eq:sourcemulti}
    \mathcal{S}_{c, i(k)} = f_\mathrm{stick}\frac{x_{v, j^\ast}\rho_\mathrm{gas}\mu_{i(k)}}{\mu_{j^\ast}\nu_{k, j^\ast}}\pi a_p^2n_pv_{j^\ast} \left(1-\frac{b_{s, i(k)}}{S_k}\right) ,
\end{equation}
where $f_\mathrm{stick}$ is the sticking probability parameter. 
In the term $(1-b_{s, i(k)}/S_k)$, $b_{s, i(k)}$ denotes the surface fraction covered by condensate $i(k)$, and $S_k$ denotes the saturation ratio of reaction $k$. 
The generalization to OM19 lies in the term $\left(1-b_{s, i(k)}/S_k\right)$, which  arises because only the surface covered with condensate $i(k)$ can evaporate, while vapor can condense onto the entire surface of the particle.
This implies that different condensates to evenly mix within the particle, which is known as the ``dirty ball'' assumption \citep{HellingWoitke2006, HellingEtal2008}.

The saturation ratio is calculated from the Gibbs energy of reactants and products and partial pressure of all reactants \citep{BerlineBricker, WoitkeEtal2018},
\begin{equation}
    \displaystyle S_i = \prod_j \left( x_{v, j}\frac{pm_\mathrm{gas}}{p^\circ \mu_j m_u} \right)^{\nu_{k, j}} \exp{\left[-\frac{\displaystyle\sum_\mathrm{product} G^\circ(T) - \sum_\mathrm{reactant} G^\circ(T)}{RT}\right]},
\end{equation}
where $p$ is the pressure, $G^\circ$ is the Gibbs formation energy measured at reference pressure $p^\circ=1\ \mathrm{bar}$, and $R$ is the ideal gas constant. We take the NIST-JANAF value for the Gibbs formation energy data \citep{Chase1998}. 
Again taking \Eq{MgSiO3} as an example, the saturation ratio of \ce{MgSiO3} amounts to
\begin{equation}
    \label{eq:saturation}
    \begin{split}
    S_{\ce{MgSiO3}} = &x_{\ce{Mg}}x_{\ce{SiO}}x_{\ce{H2O}}^2\left(\frac{p}{p^\circ}\right)^2 \\&\times\exp{\left[\frac{G^\circ_{\ce{Mg}}+G^\circ_{\ce{SiO}}+2G^\circ_{\ce{H2O}}-G^\circ_{\ce{MgSiO3}}}{RT}\right]}.\\
    =& 5.4\times 10^{37} \times x_{\ce{Mg}}x_{\ce{SiO}}x_{\ce{H2O}}^2\left(\frac{p}{p^\circ}\right)^2    (\mathrm{at}\ 1000\ K).
    \end{split}
\end{equation}
Note that in the evaluation of the above expression, $\nu_{\ce{H2}}=-2$, $x_{\ce{H2}}=1$ and $G^\circ_{\ce{H2}}=0$ by definition. The square dependence of $S_{\ce{MgSiO3}}$ on pressure comes from the summation of stoichiometric number: \ce{H2} on the product side contributes negatively to it.

\subsection{Algorithm}
\begin{table}
    \caption{Parameters that can be retrieved in \codename. In addition, \codename requires a list of cloud reactions (\Tb{reactions}) and the bulk envelope abundances (\Tb{abundance}).}
    \label{tab:parameters}
    \centering
    \small
    \begin{tabular}{c p{5.cm} l}
    \hline
    label & description & default value\\
    \hline
    \multicolumn{3}{l}{\codename PT profile parameters$^a$} \\
    $\kappa_\mathrm{IR}$ & Infrared opacity, used in T-P profile & $0.3\ \mathrm{cm}^2\mathrm{g}^{-1}$\\
    $\kappa_\mathrm{vis}$ & Visual opacity, used in T-P profile & $0.05\ \mathrm{cm}^2\mathrm{g}^{-1}$\\
    $R_\mathrm{star}$ & Stellar radius, used in T-P profile & $R_\odot$\\
    $T_\mathrm{int}$ & Internal temperature, used in T-P profile & $500\ \mathrm{K}$\\
    $T_\mathrm{star}$ & Stellar temperature, used in T-P profile & $5778\ \mathrm{K}$\\
    $a_\mathrm{pla}$ & Planet orbital radius, used in T-P profile & $0.05\ \mathrm{au}$\\
    $f_\mathrm{irr}$ & Radiation redistribution factor, used in T-P profile & $0.25$\\ 
    \hline
    \multicolumn{3}{l}{\codename cloud parameters} \\
    $\rho_n$ & Internal density of nuclei particles & $2.8\ \mathrm{g}\ \mathrm{cm}^{-3}$\\
    $\sigma_n$ & Vertical spread of nucleation region & $0.2$ \\ 
    $\dot{\Sigma}_n$ & Column integrated nucleation rate & $10^{-15}\ \mathrm{g}\ \mathrm{cm}^{-2}\ \mathrm{s}^{-1}$\\ 
    $K_{zz}$ & Diffusion parameter & $10^8\ \mathrm{cm}^2\mathrm{s}^{-1}$\\
    $P_{n}$ & Nucleation position & $60\ \mathrm{g}\ \mathrm{cm}^{-1}\ \mathrm{s}^{-2}$\\
    $a_{p0}$ & Nuclei radius & $10^{-7}\ \mathrm{cm}$\\
    $f_\mathrm{stick}$ & Sticking probability & $1$ \\
    $g$ & Surface gravitational acceleration & $2192\ \mathrm{cm}\ \mathrm{s}^{-2}$\\
    $m_\mathrm{gas}$ & Gas molecular weight & $2.34\ m_u$\\
    $x_\mathrm{v, i}$ & Vapor mass mixing ratio at the cloud base & \Tb{abundance} \\
    \hline
    \end{tabular}
    \tablefoot{
    \tablefoottext{a}{
    Note that custom T-P profile is also supported, by interpolating given temperature and pressure data. 
    }}
\end{table}

The transport Equations (\ref{eq:condensate}), (\ref{eq:vapor}) and (\ref{eq:nuclei}) are solved on a grid uniform in $\log{p}$, assuming hydrostatic balance $\frac{\partial }{\partial z} = -\frac{m_\mathrm{gas}g}{kT}\frac{\partial }{\partial \log P}$.
Our code \codename solves the boundary value problem, given by these equations through the non-linear relaxation method. The total number of equations is $n_e = n_v+n_c+1$, where $n_v$, $n_c$ are the number of vapor and condensate species considered. 
If the number of grid points is $N_\mathrm{grid}$, the total number of unknowns is $n_eN_\mathrm{grid}$.
We follow OM19 for the zero-flux upper boundary condition, $K_{zz}\frac{\partial x_{c, i}}{\partial z}-x_{c, i}v_\mathrm{sed,p}=0$ for condensate and nuclei and $\frac{\partial x_{v, j}}{\partial z}=0$ for vapor. 
In contrast to OM19, the concentration at the lower boundary is taken to be the concentration corresponding to the equilibrium condensation condition (e.g. \citet{WoitkeEtal2018}). 
This generalization alleviates the caveat of artifically setting solid concentration to  0, and allows setting the lower boundary condition at a more flexible location, e.g, within or below the cloud deck.
The computation domain is automatically adjusted to cover the entire vertical extent of the clouds. The location of the upper boundary is determined at the point where the total cloud mixing ratio is $10^{-10}$ times the peak cloud mixing ratio. The lower boundary is set at a pressure level twice the pressure where all clouds evaporate.

Above the cloud base, the particles rain down and raise their concentration at the lower layer of the atmosphere. However, they sublimate below the cloud base, causing a sharp transition from high particle concentration to cloud-free, which potentially destablizes the solution.
Therefore, care must be taken when solving for an atmosphere with strong sedimentation.

The general strategy to obtain the steady state solution is to gradually increase the level of complexity. First, we solve for equilibrium conditions before adding diffusion and settling transport to the equations.
Hence we put $K_{zz}=0$ and $v_\mathrm{sed}=0$ in Equations (\ref{eq:condensate}), (\ref{eq:vapor}) and (\ref{eq:nuclei}), which translates to $\mathcal{S}_{c,k}=0$ and $\mathcal{S}_n-x_n\rho_\mathrm{gas}/t_\mathrm{coag}=0$.
Applying Newton's method to these equations under element conservation, the mass mixing ratios ($x_{c,i}$, $x_{v,i}$ and $x_n$) found are the equilibrium solutions. Next, we gradually add diffusion by multiplying a control parameter $f_\mathrm{dif}$ to $K_{zz}$.
As $f_\mathrm{dif}$ gradually increases from 0 to 1, diffusion is added to the transport equation. Finally, the settling term is added to the equations by multiplying $v_\mathrm{sed, p}$ with $f_\mathrm{set}$, which similarly increases from 0 to 1.
The control parameters, $f_\mathrm{dif}$ and $f_\mathrm{set}$, ensure the change between each step is not too large, i.e. the solution gradually relaxes to the steady state, and successful convergence is achieved.

Accounting for particle and vapor transport (i.e. given $f_\mathrm{dif}$ and $f_\mathrm{set}$), Equations (\ref{eq:condensate}), (\ref{eq:vapor}) and (\ref{eq:nuclei}) are solved using a standard relaxation approach.
The equations can be written as a general form $\mathbf{f}(\mathbf{x})=0$, where $\mathbf{x}=(\mathbf{x}_c, \mathbf{x}_v, x_n)$. We start from a guess solution $\mathrm{x}^\mathrm{old}$ and
evaluate the residual of each equation, $\mathbf{f}(\mathbf{x}^\mathrm{old})$ on each grid point (thus $n_e\times N_\mathrm{grid}$ in total). 
Expanding the solution at the guess solution, $\mathbf{f}(\mathbf{x})\approx\mathbf{f}(\mathbf{x}^\mathrm{old})+J(\mathbf{x}^\mathrm{old})\Delta\mathbf{x}=0$, where $J(\mathbf{x}^\mathrm{old})$ is the $n_eN_\mathrm{grid}\times n_eN_\mathrm{gird}$ Jacobian matrix. Solving this linearized system, we obtain $\Delta\mathbf{x} = -J^{-1}(\mathbf{x}^\mathrm{old})\mathbf{f}(\mathbf{x}^\mathrm{old})$ and $\mathbf{x}=\mathbf{x}^\mathrm{old} + \Delta\mathbf{x}$ is used as the initial guess for next iteration, until the solution converges. 

The free parameters in the code are shown in \Tb{parameters}. We adopt the parameterized temperature-pressure (T-P) profile from \citet{Guillot2010}, which takes 7 parameters as shown by the first 7 rows in \Tb{parameters}. 
\citet{Guillot2010} follows the visual and thermal flux and calculates the temperature profile under thermal equilibrium. Besides, we note that our code also supports any other T-P profile by interpolating the user-defined T-P grid.

\subsection{Radiation Transfer}
\label{sec:RT}

\begin{table}
    \caption{Refractive index data used in this work, summarized in Table 1 of \citet{KitzmannHeng2018}}
    \label{tab:refractivedata}
    \centering
    \small
    \begin{tabular}{l l}
    \hline
    Solid & Reference \\
    \hline
    \ce{Al2O3}  &   \citet{KoikeEtal1995} \\
    \ce{Fe}, \ce{MgO}     &   \citet{Palik1991} \\
    \ce{Fe2O3} & A.H.M.J. Triaud\tablefootmark{a} \\
    \ce{Mg2SiO4}, \ce{MgSiO3}  & \citet{JagerEtal2003} \\
    \ce{FeO} & \citet{HenningEtal1995} \\
    \ce{FeS} & \citet{PollackEtal1994} \\ 
    \ce{KCl}, \ce{NaCl} & \citet{Palik1985} \\
    \ce{Na2S} & \citet{KhachaiEtal2009} \\
    \hline
    \end{tabular}
    \tablefoot{
    \tablefoottext{a}{\url{https://www.astro.uni-jena.de/Laboratory/OCDB/mgfeoxides.html}}
    }
\end{table}

Solid opacities originating from cloud particles are of vital importance in determining transmission or emission spectra. 
Effective medium theory averages the (complex) refractive index $m_i$ of each component constituting the ``dirty cloud particle'', given their volume mixing ratio $f_V$. Bruggeman theory (see, for example, \citet{Choy2016}) assuming randomly mixed spherical monomer made of homogeneous material, gives the expression from which the averaged refractive index $m$ can be solved:
\begin{equation}
    \label{eq:B-mixing}
    \sum_{i} f_{V,i} \frac{m_i^2-m^2}{m_i^2+2m^2} = 0 ,
\end{equation}
where the refractive indices of each pure material, $m_i$, has been compiled by \citet{KitzmannHeng2018} (\Tb{refractivedata}). 
The Bruggeman mixing rule is widely adopted in the literature \citep{MishchenkoEtal2016, LinesEtal2018, MinEtal2020}, but does rely on assumptions that may in certain situations be less appropriate. In particular, while we use the DHS model for the shape of the particles (next paragraph), the mixing rule in \eq{B-mixing} assumes that each component within the particle is made up of tiny spherical constituents, homogeneously mixed throughout the particle. The spherical nature is sometimes problematic as the effective dielectric constant may be different from other shapes (e.g., Fe needles are far more opaque than Fe spherules). Recently, \citet{KieferEtal2024b} have compared how different opacity mixing treatments affect the spectra, including different mixing rules and core-shell particle structure. It still remains an open question what the morphology of a heterogeneous cloud particle is in reality and which opacity treatment to take.

Besides the material refractive index, cloud particles' shape and size also determine their absorption and scattering properties. We assume the particle shape follows the distribution of hollow spheres (DHS) \citep{MinEtal2005}, because, on average, the optical properties of DHS can mimic an ensemble of irregularly shaped particles.
\citet{MinEtal2005} and \citet{MolliereEtal2017} shows that the scattering properties of DHS is different from spherical particles whose scattering properties are given by Mie theory at mid-IR wavelength. Based on DHS, we use software \texttt{OpTool} \citep{DominikEtal2021} to calculate the particle scattering and absorption opacity.

Apart from clouds, gas molecules also contribute to the atmosphere opacity, with their spectral lines indicative of the atmospheric component. 
The gas opacity in our model includes all the vapor species involved in cloud-formation (\Tb{reactions}), plus other gas molecules abundant in exoplanet atmosphere -- \ce{CO}, \ce{CO2} and \ce{CH4}. We also consider Rayleigh scattering of \ce{H2}, \ce{He}, and collision-induced absorption (\ce{H2}-\ce{H2} and \ce{H2}-\ce{He}).
After cloud formation we forced the gas phase concentration to local thermo-chemical equilibrium values. This is done by mixing the ``leftover vapor'' and background gas (\ce{H2}, \ce{He}, \ce{CO}, \ce{CO2} and \ce{CH4}) locally and running \texttt{FastChem} \citep{StockEtal2022} to compute gas composition assuming locally chemical equilibrium. 
How disequilibrium chemistry like photoevaporation and diffusive quenching \citep{MadhusudhanSeager2011, TsaiEtal2017, KieferEtal2024} affects the spectrum, is beyond the scope of this paper.
Finally \texttt{petitRADTRANS} \citep{MolliereEtal2019} is run to calculate the transmission and emission spectra, based on the equilibrium gas concentration and the cloud opacity described in the above paragraph. The reference pressure in \texttt{petitRADTRANS} is set to $10^{-4}$ bar for all cases.

\section{Results} \label{sec:result}

\begin{figure}
    \includegraphics[width=\columnwidth]{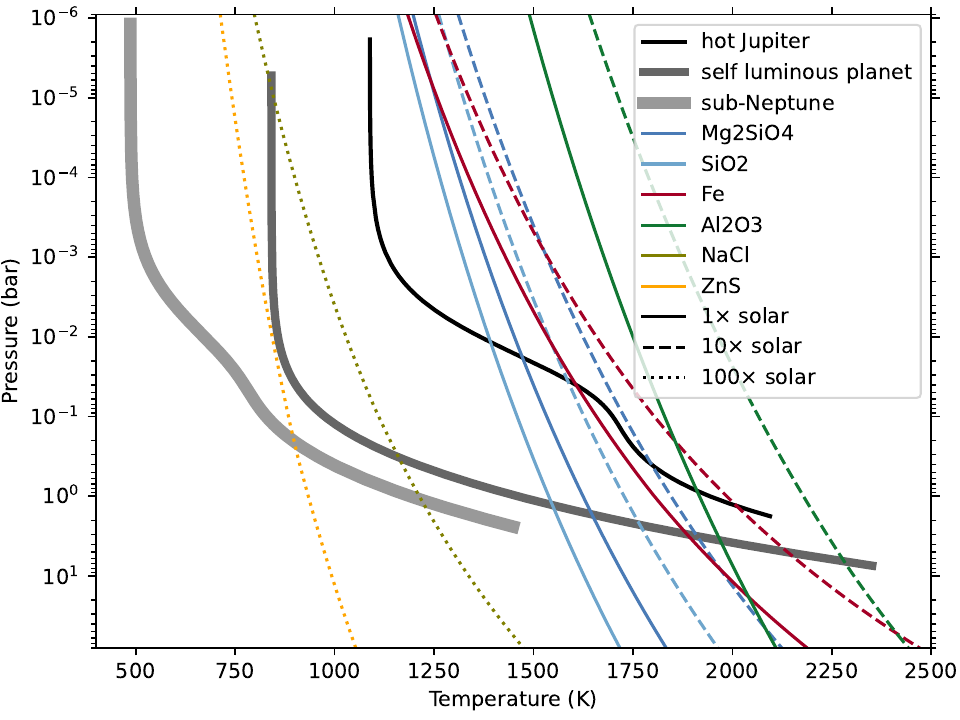}
    \caption{\label{fig:TP} 
    Default temperature-pressure profile of the hot Jupiter (\Se{result}), self-luminous (\Se{HR8799}) and sub-Neptune planet (\Se{GJ1214}), following \citet{Guillot2010}. The irradiation temperatures of the hot Jupiter, self-luminous planet and the sub-Neptune planet are taken to be $500\ \mathrm{K}$, $1000\ \mathrm{K}$ and $300\ \mathrm{K}$, correspondingly. 
    Condensation curves of \ce{Mg2SiO4}, \ce{SiO2}, \ce{Fe}, \ce{Al2O3} (for hot Jupiter and self-luminous planet) and \ce{NaCl} and \ce{ZnS} (for sub-Neptune) are shown as colored lines. The solid, dashed and dotted saturation curves denote solar, 10 times solar and 100 times solar metallicity, respectively.}
\end{figure}

\begin{table}
    \caption{The cloud-forming reactions considered in the hot Jupiter model and sub-Neptune model.}
    \label{tab:reactions}
    \setlength{\tabcolsep}{0.3pt}
    \centering
    \small
    \begin{tabular}{c | r c l}
    \hline
    & \ce{Mg + SiO + 2H2O} & \ce{->} & \ce{MgSiO3(s) + 2H2}\\
    & \ce{2Mg + SiO + 3H2O} & \ce{->} & \ce{Mg2SiO4(s) + 3H2}\\
    & \ce{SiO + H2O} & \ce{->} & \ce{SiO2(s) + H2}\\
    & \ce{Mg + H2O} & \ce{->} & \ce{MgO(s) + H2}\\
    hot & \ce{Fe + H2O} & \ce{->} & \ce{FeO(s) + H2}\\
    Jupiter & \ce{Fe + H2S} & \ce{->} & \ce{FeS(s) + H2}\\
    & \ce{2Fe + 3H2O} & \ce{->} & \ce{Fe2O3(s) + 3H2}\\
    & \ce{Fe} & \ce{->} & \ce{Fe(s)}\\
    & \ce{TiO + H2O} & \ce{->} & \ce{TiO2(s) + H2}\\
    & \ce{2Al + 3H2O} & \ce{->} & \ce{Al2O3(s) + 3H2}\\
    \hline
    & \ce{2Na + 2HCl} & \ce{->} & \ce{2NaCl(s) + H2}\\
    sub- & \ce{2K + 2HCl} & \ce{->} & \ce{2KCl(s) + H2}\\
    Neptune & \ce{2Na + H2S} & \ce{->} & \ce{Na2S(s) + H2}\\
    & \ce{Zn + H2S} & \ce{->} & \ce{ZnS(s) + H2}\\
    \hline
    \end{tabular}
\end{table}

\begin{table}
    \caption{Gas species mass mixing ratio at the lower boundary in the hot Jupiter models. The default model assumes a solar metallicity and C/O. In $10\times$-[Z] model, the metallicity is elevated by an order of magnitude above solar. 
    In C/O$=0.9$ model, the C-to-O number ratio is changed to 0.9. Note that CO gas is non-condensing, but its abundance affects the transmission spectrum.}
    \label{tab:abundance}
    \centering
    \small
    \begin{tabular}{c c c c}
    \hline
    gas & default & $10\times$-[Z] & C/O$=0.9$\\
    \hline
    \ce{Mg} & $4.1\times 10^{-4}$ & $4.1\times 10^{-3}$ & $4.1\times 10^{-4}$\\
    \ce{SiO} & $6.1\times 10^{-4}$ & $6.1\times 10^{-3}$ & $6.1\times 10^{-4}$\\
    \ce{H2O} & $1.4\times 10^{-3}$ & $1.4\times 10^{-2}$ & $6.8\times 10^{-5}$\\
    \ce{Fe} & $7.6\times 10^{-4}$ & $7.6\times 10^{-3}$ & $7.6\times 10^{-4}$\\
    \ce{H2S} & $1.9\times 10^{-4}$ & $1.9\times 10^{-3}$ & $1.9\times 10^{-4}$\\
    \ce{TiO} & $2.4\times 10^{-6}$ & $2.4\times 10^{-5}$ & $2.4\times 10^{-6}$\\
    \ce{Al} & $3.2\times 10^{-5}$ & $3.2\times 10^{-4}$ & $3.2\times 10^{-5}$\\
    \hline
    \ce{CO} & $3.2\times 10^{-3}$ & $3.2\times 10^{-2}$ & $4.4\times 10^{-3}$\\
    \ce{K} & $1.7\times 10^{-6}$ & $1.7\times 10^{-5}$ & $1.7\times 10^{-6}$\\
    \ce{Na} & $1.7\times 10^{-5}$ & $1.7\times 10^{-4}$ & $1.7\times 10^{-5}$\\
    \hline
    \end{tabular}
\end{table}

Similar to OM19, our default model investigates a generic hot Jupiter at a distance of 0.05 au from a solar-like star. The temperature is prescribed by \citet{Guillot2010} accordingly (see \Fg{TP}), with internal heat flux corresponding to $T_\mathrm{int}=500\,\mathrm{K}$.
Taking a diffusivity of $K_{zz}=10^8\ \mathrm{cm}^2 \mathrm{s}^{-1}$ and solar elemental abundance, we run \codename to investigate the cloud forming on the above hot Jupiter. The envelope bulk elemental abundances at the lower boundary is listed in the first column of \Tb{abundance}, and the reactions we consider in the default model are listed in \Tb{reactions}.
In \Se{cloudsolid} and \ref{sec:cloudgas}, we present the results of the default model, focusing on solid and gas, respectively. In \Se{comparison}, we compare \codename with other cloud formation models. 
Then, in \Se{parameterstudy}, we discuss how $K_{zz}$, the nucleation profile, the choice of cloud species, and the molecular abundance at the boundary condition influences the cloud formation.
Following that, in \Se{spectrum}, radiation transfer calculations are performed to investigate how the gas and cloud species influence the transmission spectrum. 

\subsection{Cloud Structure}
\label{sec:cloudsolid}

\begin{figure*}
    \sidecaption
    \includegraphics[width=0.7\textwidth]{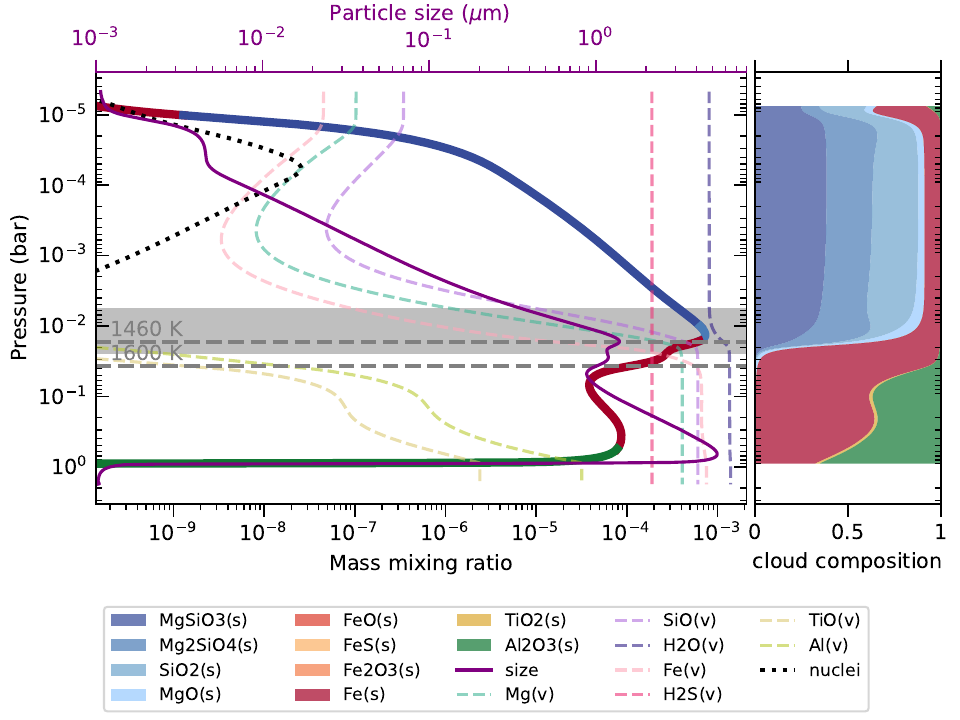}
    \caption{\label{fig:default} 
    Atmosphere structure of the default hot Jupiter model. The thick colored line shows the cloud mass mixing ratio summed over all species. The dominant species by mass is illustrated by the color of the thick line. Vapor mixing ratios are shown by dashed lines.
    The dotted black line denotes the nuclei mass mixing ratio and the purple thin line denotes the cloud particle size, which should be referred to the upper axis. In the figure we also labelled the evaporation temperature of magnesium-silicates (1460 K) and iron (1600 K) with horizontal dashed lines. 
    The grey band refers to where the cloud optical depth in the IR band ($1\ \mathrm{\mu m}-20\ \mathrm{\mu m}$), integrated from the top of the atmosphere, reaches 1. 
    The right panel shows the composition of cloud particles in mass.}
\end{figure*}

\Fg{default} presents the solid, gas, nuclei mixing ratios and particle size in the default cloud model. The total column density of the cloud amounts to $0.03\ \mathrm{g}\ \mathrm{cm}^{-2}$. The cloud shows a two-layer structure: at around $10^{-2}\ \mathrm{bar}$, it is dominated by \ce{MgSiO3}, \ce{Mg2SiO4} and \ce{SiO2} (refered to as magnesium-silicate or Mg-Si hereafter), whereas beyond $2\times 10^{-2}\ \mathrm{bar}$, \ce{Fe}-dominant condensates take over. 
The layered structure is consistent with the condensation sequence of magnesium-silicate (${\approx}1460\,\mathrm{K}$) and iron condensate (${\approx}1600\,\mathrm{K}$) species, respectively. 
As a result, the magnesium-silicate condensate group evaporates higher in the atmosphere than Fe.
There is no cloud below 1 bar, because even the most refractory species considered here, \ce{Al2O3}, evaporates. 
The right panel of \Fg{default} gives a graphical illustration of the components making up the cloud particles by mass. At the location of the magnesium-silicate cloud ($P<0.02\ \mathrm{bar}$), though the temperature is below the iron condensation point, the cloud particles are depleted of iron, because the iron element settles to deeper atmosphere.
In the iron-dominated regions of the cloud, at around $0.2\ \mathrm{bar}$, we still find $1\%$ of \ce{MgO} and \ce{SiO2}, even though equilibrium chemistry would predict their absence due to the temperature higher than the magnesium-silicates evaporation value. 
This result arises from the rapid downward transport by settling and the ``dirty ball'' assumption: the magnesium-silicates are transported to the level due to settling and they do not sublimate efficiently limited by the surface coverage.
Due to settling, the condensate mass mixing ratio increases with depth and peaks near the bottom of the cloud (${\approx} 10^{-2}\ \mathrm{bar}$ for magnesium-silicates and ${\approx} 0.2\ \mathrm{bar}$ for Fe condensates).
During settling, coagulation and condensation grow the particle larger, reaching $4\ \mu\mathrm{m}$ at $0.5 \ \mathrm{bar}$ (\Fg{default}) and leading to faster settling at lower levels. 
As the temperature increases with pressure, the cloud quickly evaporates at the bottom, causing a steeper  slope at the cloud bottom than the cloud top.

Nuclei mass mixing ratio $x_n$, which is proportional to particle number density, is subject to both production and coagulation.
The nuclei mixing ratio peaks at a height corresponding to $6\times 10^{-5}\ \mathrm{bar}$ (see \Fg{default}), where they are injected in the default simulation. From their formation location the particles diffuse up and settle down, leading to decreasing particle number density, towards both upper and lower atmosphere. 
Particles hardly grow beyond their initial size ($1\ \mathrm{nm}$) at the location where the nuclei are injected, because they diffuse away before coagulate. 
As the nuclei mixing ratio decreases towards higher pressure, the particle size generally increases, despite a peak at $10^{-2}\ \mathrm{bar}$, which is due to the magnesium-silicate cloud.

\subsection{Molecular Gas Profile}
\label{sec:cloudgas}

The vapor mass mixing ratios are shown by dashed lines in \Fg{default}. Near the cloud base, all vapor species approach their boundary abundances, due to the evaporation of all cloud species. 
\ce{Fe}, \ce{TiO}, \ce{Al}, \ce{SiO} and \ce{Mg} are significantly depleted in the upper layers, because they are the limiting species that form \ce{Fe}(s), \ce{TiO2}, \ce{Al2O3} and magnesium-silicates, respectively. 
The cloud formation (condensation) results in a minimum region in the vapor cencentration (e.g. $4\times 10^{-4}\ \mathrm{bar}$ for \ce{H2O}), to which vapor from lower regions diffuses. Above $p=4\times 10^{-4}\ \mathrm{bar}$, vapor mixing ratio rises again with atmospheric height. 
The vapor here comes from the evaporation of the cloud particles, due to lower pressure.
The vapor mass mixing ratio approaches a constant value at the upper boundary, because the diffusive flux is imposed to be 0. In addition, non-limiting vapor species also deplete, albeit less prominently. For example, \ce{H2O} vapor mass mixing ratio decreases by a factor of 2 in the upper atmosphere.
Therefore, consumption of condensable vapor due to cloud formation may result in less distinct molecular lines than the cloud-free case. 
This quenching effect implies that  assuming constant gas abundance during retrieval may underestimate the  metallicity of an exoplanet. The C/O ratio of the atmosphere is also affected by the formation of cloud. As the cloud formation consumes O-bearing vapor like \ce{H2O}, consequently the atmospheric C/O ratio increases from 0.52 to 0.54 

\subsection{Comparison with Previous Studies}
\label{sec:comparison}
\begin{figure*}
    \includegraphics[width=\textwidth]{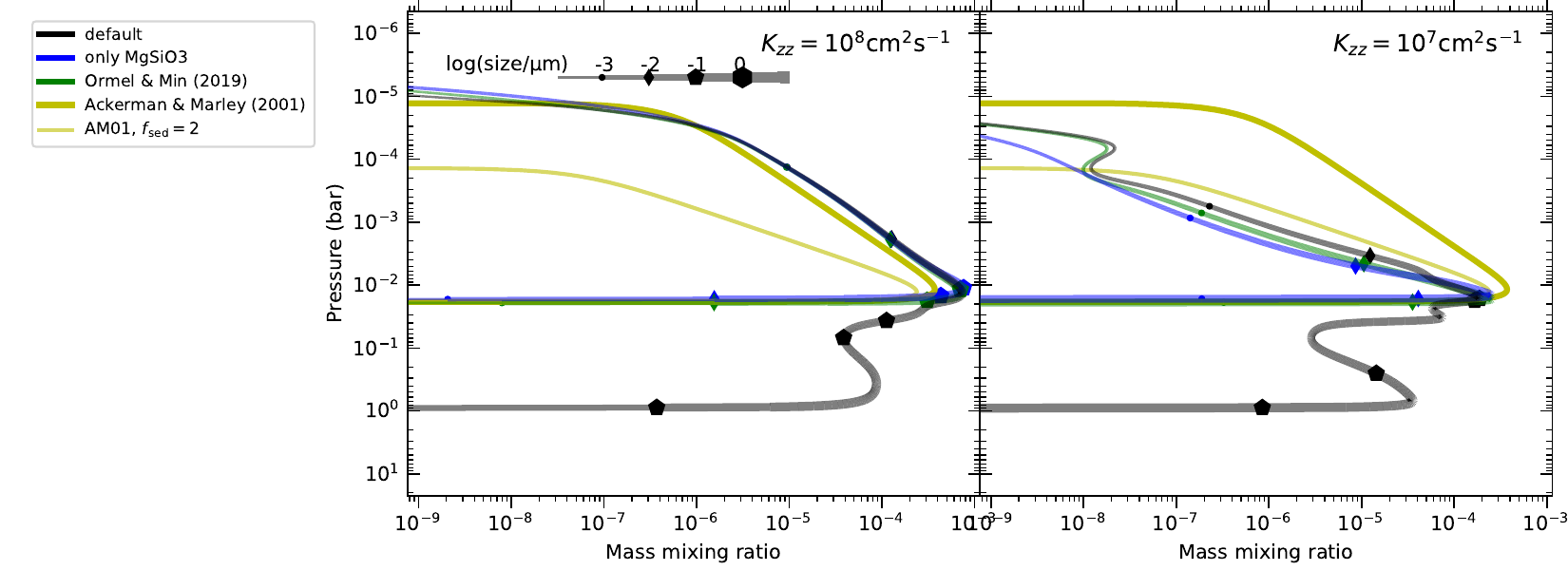}
    \caption{\label{fig:comparison}
    Comparison of the full \codename model (default), \codename restricted with \ce{MgSiO3} condensates only, the \citet{OrmelMin2019} model, and \citet{AckermanMarley2001}. The thickness of the lines and the symbols on the lines indicates the particle size, except in the \citet{AckermanMarley2001} runs, where the particle size is not solved for as it decouples with the cloud concentration in the model.
    \textbf{Right}: the same as left, but $K_{zz}$ decreased to $10^7\ \mathrm{cm}^2\mathrm{s}^{-1}$. In this case, the $f_\mathrm{sed}=2$ run (thin line) provides a better correspondence to other models.
    }
\end{figure*}

In this section we compare \codename with \citet{AckermanMarley2001} (AM01) and OM19. 
The main advantage of \codename lies in the self-consistent formation of multiple cloud species, compared to the single species cloud in AM01 and OM19. \citet{MinEtal2020} extended the OM19 method, but the multi-species cloud formation sequence is prescribed.
For a meaningful comparison, we only include \ce{MgSiO3} condensate in three models. Assuming equilibrium between sedimentation and diffusion, AM01 solves the equation for $x_t$, the total molecular concentration in gas and solid phase,
\begin{equation}
    \label{eq:AM01}
    K_{zz}\frac{\partial x_t}{\partial z} + f_\mathrm{sed}w_{\ast}x_c = 0,
\end{equation}
where $w_\ast$ is the convective velocity scale, and $f_\mathrm{sed}$ is an order of unity settling coefficient. We take default $f_\mathrm{sed}=1$, $w_\ast = K_{zz}/H$, where $H$ is the scale height of the atmosphere, and 
\begin{equation}
    x_c = \max{\{0, x_t-x_s\}}, \label{eq:AMxc}
\end{equation}
where $x_s$ is the saturation mixing ratio of \ce{MgSiO3}. The \ce{MgSiO3} saturation vapor pressure in both AM01 and OM19 is $P_s = 1.04\times 10^{17} \exp{\left[58663/T\right]}\ \mathrm{dyn\ m^{-2}}$ \citep{AckermanMarley2001}.
After $x_t$ is calculated, cloud condensate mixing ratio $x_c$ can be deduced through \Eq{AMxc}.

The results are shown in \Fg{comparison}.  Generally a good match is seen between the three models. For a single species cloud, the transport equations we solve are the same as those in OM19, except the expression for saturation ratio and condensation rate (\Fg{cartoon}). As the saturation pressure used in AM01 and OM19 can translate to a reasonable estimation to the saturation ratio calculated from a more formal way (\Eq{saturation}), OM19 predicts a similar cloud profile to ours. 
However, we warn that unlike OM19's assumption, the precise saturation ratio depends not only on temperature, but also on the molecular abundance (\Eq{saturation}), e.g. metallicity or C/O ratio. 
Therefore, the saturation pressure of \ce{MgSiO3} adopted in OM19 cannot be universally applied.

The settling velocity in AM01 model is characterized by the $f_\mathrm{sed}$ parameter, rather than calculated from physical principles. Though the default AM01 run assuming $f_\mathrm{sed}=1$ agrees with our model reasonably well, it generally underestimates the cloud mixing ratio by a factor of 2. Furthermore, if $f_\mathrm{sed}$ is raised to 2 (thin yellow line), the extent of the cloud in AM01 shrinks significantly and no longer matches our model. Therefore, the overall agreement between AM01 and our model seems to be coincidental in the sense that the settling velocity corresponding to $f_\mathrm{sed}=1$ happens to be similar to the settling velocity of $a\sim 10^{-2}\ \mathrm{\mu m}$ particle.
This can be tested through changing $K_{zz}$ and other parameters. In particular, the AM01 cloud is independent of both $K_{zz}$, which cancels in \Eq{AM01} through the choice of $w^\ast$, and the nucleation profile. In the right panel of \Fg{comparison}, $f_\mathrm{sed}=2$ provides a better estimation of the $K_{zz}=10^7\ \mathrm{cm}^2 \mathrm{s}^{-1}$ cloud than $f_\mathrm{sed}=1$. However, as we will see in \Se{parameterstudy}, $K_{zz}$ greatly affects the cloud thickness and concentration.

\subsection{Parameter Study}
\label{sec:parameterstudy}

\begin{figure*}
    \includegraphics[width=\textwidth]{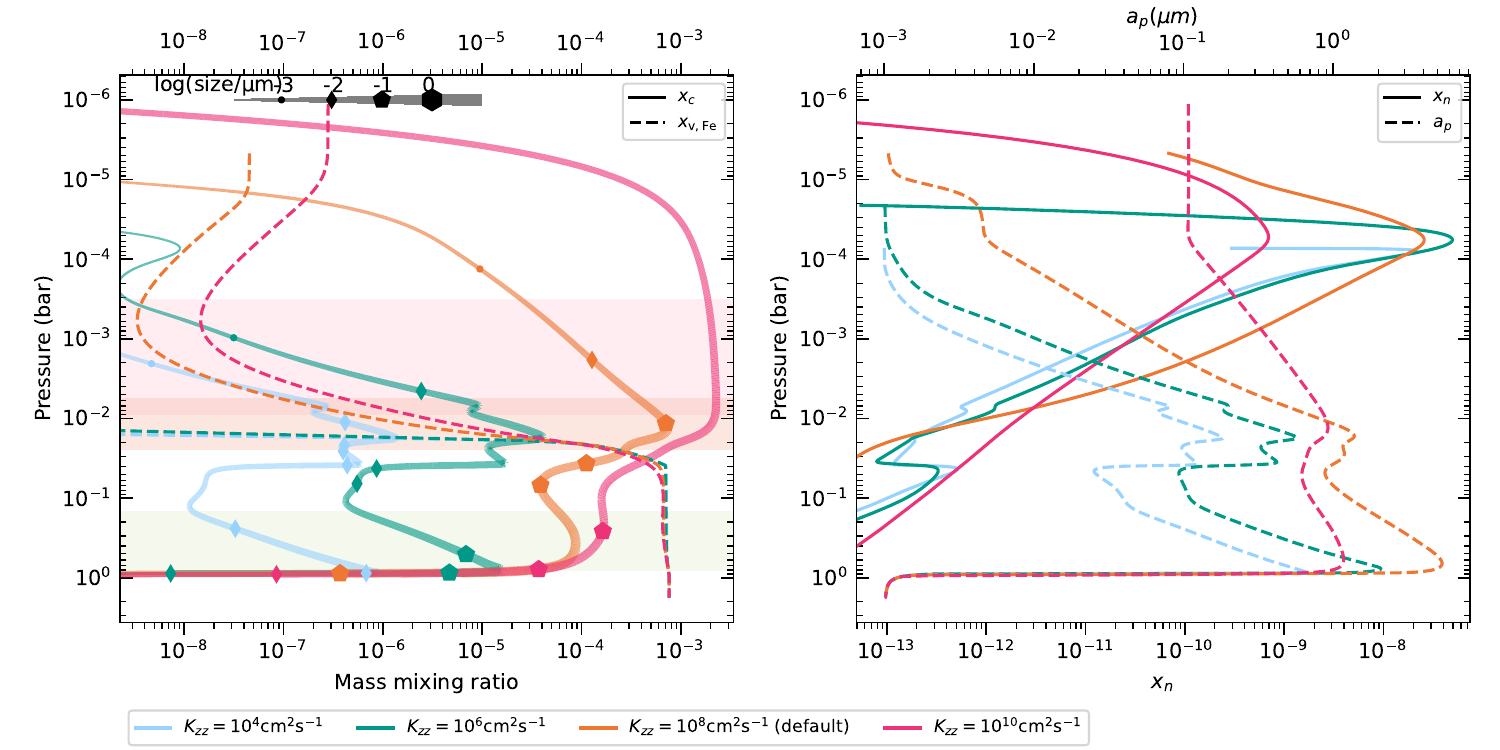}
    \caption{\label{fig:paraKzz}
    Cloud profiles under different eddy diffusivity parameters $K_{zz}$, which serve as effective vertical turbulence. Left: total cloud mixing ratio (solid lines) and \ce{Fe} vapor mixing ratio (dashed lines) as an example. Right: nuclei mixing ratio (solid lines) and particle size (dashed lines). Similar to \Fg{default}, the location where the cloud optical depth reaches unity is shown in the left panel by the horizontal bands with corresponding colors.
    Note that there is no blue band corresponding to $K_{zz} = 10^4\ \mathrm{cm^2s^{-1}}$, because the cloud optical depth never reaches 1.}
\end{figure*}

\subsubsection{Atmosphere diffusivity}
The eddy diffusivity $K_{zz}$ determines the extent of cloud, because the vertical turbulent motion of gas serves as an effective diffusivity term, blowing up the trace gas and solid. Formally $K_{zz}$ is derived from mixing length theory \citep[e.g.][]{WangEtal2015} or from hydrodynamical simulations by measuring the mixing flux and calculate how it correlates with the concentration gradient. (e.g. \citet{ZhangShowman}). 
$K_{zz}$ depends on the atmospheric thermal structure, heating flux, precipitation and phase change, as is suggested by \citet{GeEtal2023, LeconteEtal2024}. Due to the uncertainty in $K_{zz}$, in our study we use a constant $K_{zz}$ as a representative value over the entire atmosphere.

We test how the choice of $K_{zz}$ influence the cloud in \Fg{paraKzz}. The effect of increasing $K_{zz}$ is twofold: (i) cloud particles are transported upwards -- counter-acting settling; (ii) the cloud-forming vapor is replenished more efficiently. As a result, the total cloud mixing ratio increases with $K_{zz}$. A box-like cloud, extending from $10^{-6}\ \mathrm{bar}$ to $1\ \mathrm{bar}$, appears for the $K_{zz}=10^{10}\ \mathrm{cm^2 s^{-1}}$ run. 
In contrast, the $K_{zz}=10^4\ \mathrm{cm^2 s^{-1}}$ or $K_{zz}=10^6\ \mathrm{cm^2 s^{-1}}$ runs exhibit a clear two-layered cloud structure, because settling is more effective. 
As a result of the thinner cloud layer, the cloud optical depth $\tau_\mathrm{cloud}=1$ surface lies deeper in the atmosphere for lower $K_{zz}$. In the case of very low diffusivity, $K_{zz}=10^4\ \mathrm{cm^2 s^{-1}}$, the cloud opacity never reaches unity. 

Comparing the $K_{zz}=10^{6}\ \mathrm{cm^2 s^{-1}}$ to $K_{zz}=10^{4}\ \mathrm{cm^2 s^{-1}}$ case, the particle number density roughly overlaps, because coagulation dominates the particle number density distribution when the transport is weak. As a result, the particle size increases mainly due to the increasing amount of condensates. When $K_{zz}$ increases to $10^{8}\ \mathrm{cm^2 s^{-1}}$, more particles are transported downward up to $10^{-2}\ \mathrm{bar}$ due to stronger diffusion. The particles grow so large that they settle quickly near the cloud bottom (${<}10^{-2}\ \mathrm{bar}$). Therefore, the particle nuclei near the cloud bottom deplete and in turn raise the particle sizes to $4\ \mathrm{\mu m}$. At the other extreme, for $K_{zz}>10^{10}\ \mathrm{cm^2 s^{-1}}$ transport is dominated by diffusion, and the particle number density at the cloud lower boundary increases with stronger diffusive supply. Consequently, the particle size at the cloud bottom decreases when the diffusion is strong.
For $K_{zz}=10^{10}\ \mathrm{cm}^2\mathrm{s}^{-1}$, the particle size reaches $0.1\ \mathrm{\mu \mathrm{m}}$ throughout the cloud. 

We also plot the \ce{Fe} mixing ratio to examine how vapor content is affected by $K_{zz}$. With stronger diffusion, there is more \ce{Fe} vapor diffused up, although its concentration above the cloud formation layer is negligible in all $K_{zz}$ runs. Our result implies that in a strongly turbulent atmosphere, the molecular features will become less apparent, or even indiscernible, given the thicker cloud. 

\begin{figure}
    \includegraphics[width=\columnwidth]{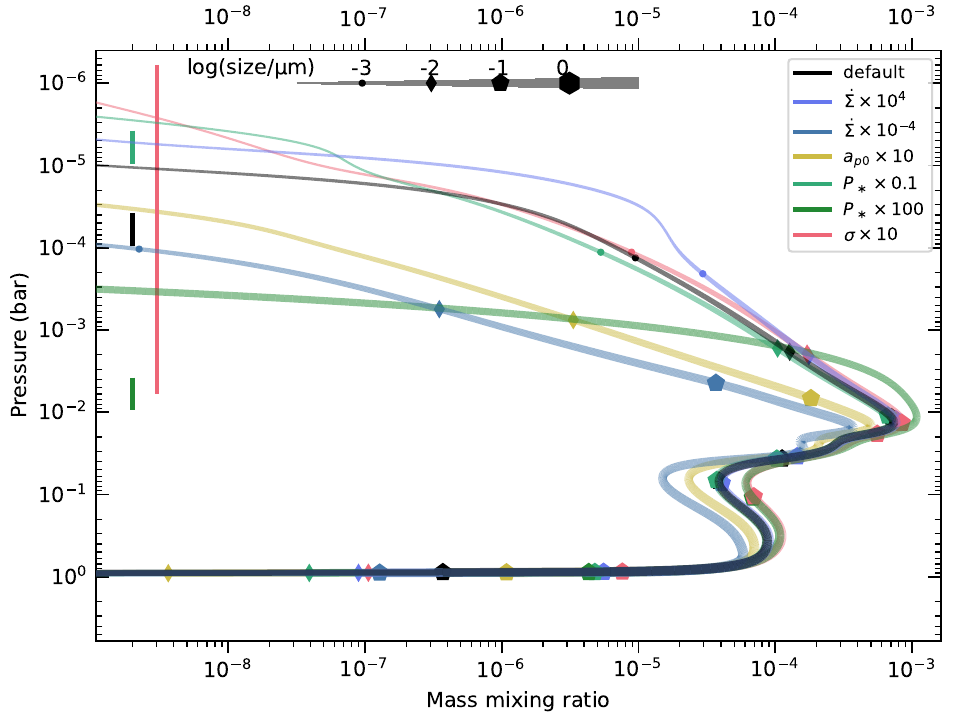}
    \caption{\label{fig:paranuc}
    Sensitivity of the nucleation profiles on the cloud profiles. In the figure we changed the nucleation strength $\dot{\Sigma}_n$, nucleation location $P_n$, spread of nucleation $\sigma_n$ and nuclei size $a_{p0}$. 
    The parameters are changed with respect to the default value in \Tb{parameters}. The vertical lines indicate the location of nucleation, with its length meaning $1\sigma$ dispersion of the nucleation profile.}
\end{figure}

\subsubsection{Nucleation}\label{sec:nucleation}

As discussed in \Se{singlemethod}, how much and where nuclei form on exoplanets is a controversial topic of atmosphere modelling. In \codename, the profile of the nucleation production rate is entirely parameterized. We find that the nucleation profile plays a less important role in affecting the cloud than eddy diffusivity. In \Fg{paranuc} we vary the imposed nucleation profile, including nucleation strength $\dot{\Sigma}_n$, nucleation position $P_n$, nucleation spread $\sigma_n$ and nuclei particle size $a_{p0}$.
Increasing the nucleation rate by 4 orders of magnitude has little effect on the cloud structure. This is because the amount of cloud condensation primarily depends on the vapor concentration, rather than the amount of nuclei. When there are ample amount of nuclei, coagulation among particles mitigates the effect of further increasing the nuclei injection rate. On the other hand, when fewer nuclei are injected, the cloud particles’ sizes grow larger, resulting in a more compact cloud that lies deeper in the atmosphere. 

Changing the nucleation location hardly influence the clouds below it (light and dark green line in \Fg{paranuc}). This is because nuclei formed higher in the atmosphere can settle down to lower levels. 
For the same reason, the cloud profile changes little when there is larger spread in the nucleation (the pink line). 
However, when nuclei are injected deeper in the atmosphere, the cloud will not extend much upwards because the particles have grown too large to reach higher regions by diffusing up against gravity, leading to less nuclei concentration above the nucleation pressure.
We do observe an increase of particle size at around $10^{-3}\ \mathrm{bar}$ and a reduction of the cloud extent when nucleation proceeds at a pressure scale that is two orders of magnitude higher than in the default model.
In hot Jupiters, $({\ce{TiO2}})_N$ clustering can be a major source of nucleation. Deriving updated Gibbs energy of $({\ce{TiO2}})_N$ clusters, \citet{LeeEtal2015} found \ce{TiO2} nucleation happens around a few $10^{-4}\ \mathrm{bar}$, broadly consistent with our fiducial choice of the nucleation profile.
Photochemistry haze, another potential source of nucleation on hot Jupiter's atmosphere, is found to form around $10^{-5}$ to $10^{-6}\ \mathrm{bar}$ \citep{ArfauxLavvas2022, LavvasKoskinen2017, ArfauxLavvas2024}. Therefore, though the actual location of nucleation on a hot Jupiter is uncertain, our fiducial clouds result can represent the real case.

Lastly, if the nuclei initial size is increased to $10\ \mathrm{nm}$, the upper cloud extent is diminished. The effect is similar to decreasing $\dot{\Sigma}_n$, which also acts to increase the cloud particles' sizes. A larger nuclei size is more appropriate if the Kelvin effect is strong. 
The Kelvin effect expresses the suppression of condensation due to the curvature of the nuclei. When the particle size ($a$) is small and the surface tension of the material ($\Gamma$) is high, it reduces the condensation exponentially by a factor 
\begin{equation}
\label{eq:Kelvin}
\exp{\left(\frac{2\Gamma \rho_\bullet \mu m_u}{k_b T a}\right)}
\end{equation}
where $\mu m_u$ is the molecular weight and $T$ is the temperature. To overcome the Kelvin effect, nuclei are likely to coagulate to a larger radius, an effect that can be modeled in \codename by choosing a larger size for the nuclei.
In summary, although \codename cloud results are insensitive to nucleation, assumptions for large nucleation rates, its nucleation parameters provide the flexibility to mimic the real nucleation profile.

\begin{figure}
    \includegraphics[width=\columnwidth]{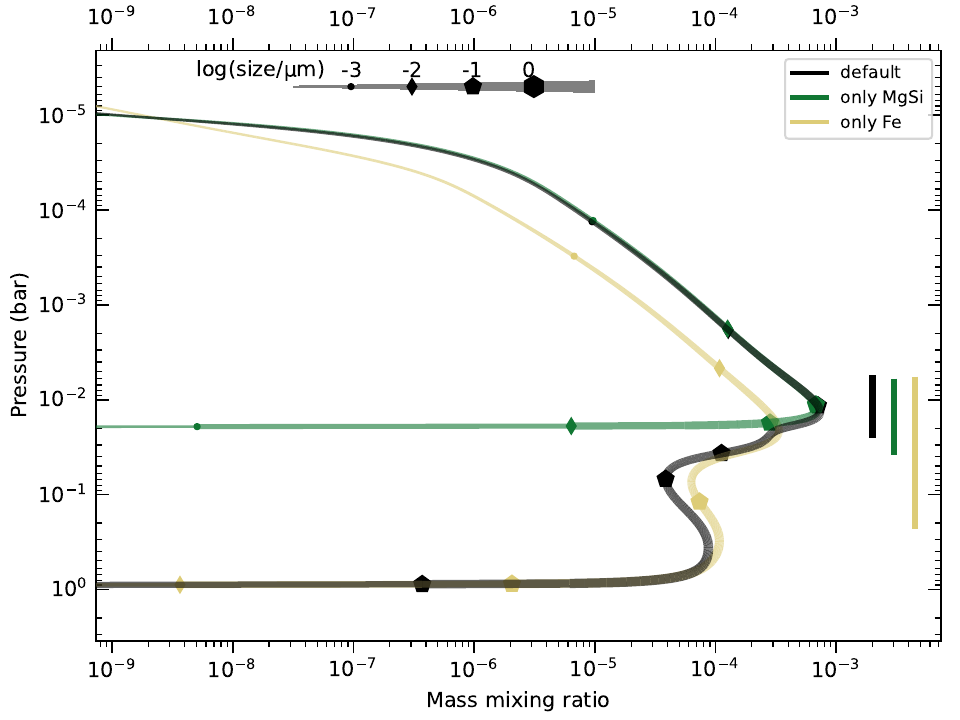}
    \caption{\label{fig:pararec}
    Cloud profiles with different condensation reactions. The full set of reactions \Tb{reactions} is shown as the black line. The green and yellow line only includes the magnesium-silicate group versus the iron and refractory reaction group, i.e. the first four rows and the last six rows in \Tb{reactions}. The vertical lines represent where the near-IR optical depth accumulates to unity. 
    }
\end{figure}

\subsubsection{Condensation reactions}

\Fg{pararec} illustrates how the included condensation reactions affect the physical properties of the cloud. For ``only MgSi'', we only include the first 4 reactions in \Tb{reactions}, whereas for ``only Fe'' we only include the last 6 reactions, which is representative of the more refractory cloud species. 
With ``only MgSi'' reactions, one can recover the entire cloud mixing ratio above the magnesium silicate evaporation line, while with ``only Fe'' model one can recover the iron and refractory peaks at $10^{-2}$ to $1$ bar. 
Compared to the full model (\Fg{default}), in the ``only Fe'' model, the iron and refractory cloud also occupies the height dominated by magnesium-silicate cloud in the default model. 
Though \ce{Fe} is more opaque than MgSi, the $\tau=1$ of ``only Fe'' model is similar to that of the default model, because the \ce{Fe} particles are smaller.
As the $\tau=1$ surface (given by the vertical lines on the right) lies roughly in the magnesium silicate cloud, only including magnesium silicates condensates is still an acceptable estimation to the full reactions. However, this may not hold when the cloud optical depth would be lower, e.g., when $K_{zz} \lesssim 10^6\ \mathrm{cm^2 s^{-1}}$.

\begin{figure*}
    \includegraphics[width=\textwidth]{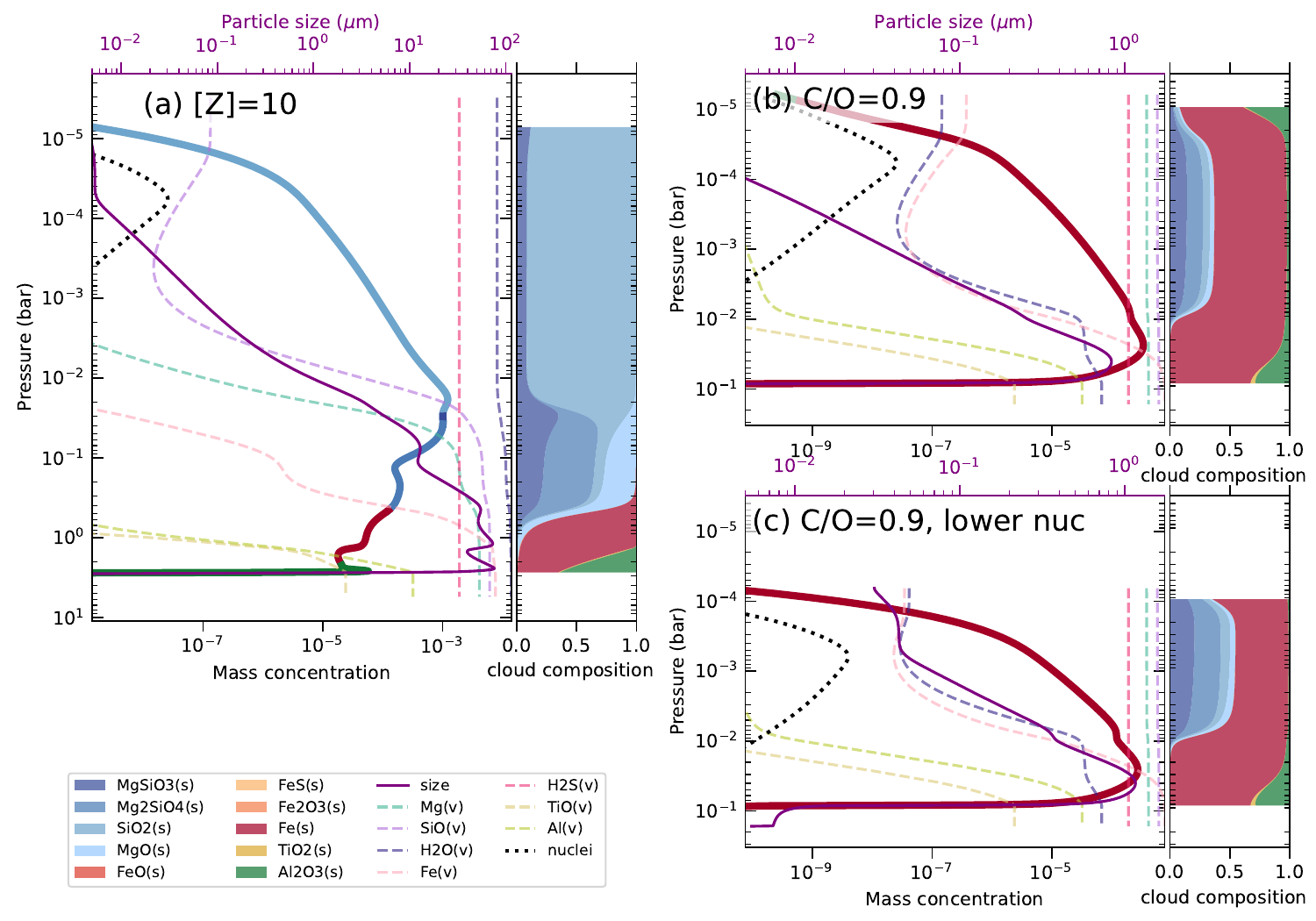}
    \caption{\label{fig:paraboundary}
    Effect of vapor abundances at the lower boundary of atmosphere on the solid, vapor, nuclei profile, particle size and cloud composition. (a) 10$\times$-{Z} hot Jupiter model; (b) C/O=0.9 hot Jupiter model; (c) C/O=0.9 hot Jupiter model, like (b), but with injection of larger nuclei deeper in the atmosphere, mimicking nucleation featuring high particle surface tension. The thick solid, dashed and dotted lines indicate total cloud, vapor and nuclei mixing ratios, respectively. The purple solid line shows the particle size. The right sub-panel in each panel shows the composition of the cloud particles in mass.  
    Compared to the default model (\Fg{default}) the upper cloud layers of the $10\times$-[Z] run become devoid in \ce{Mg} (\ce{SiO2} dominates) while in the C/O=0.9 run \ce{Fe}-bearing cloud particles make it into the upper regions.
    }
\end{figure*}

\subsubsection{The effect of envelope bulk abundance on the cloud properties}\label{sec:bulk}

For the default model, we assume solar values \citep{AsplundEtal2009} for the planet's envelope elemental abundances. Based on that, we assume that \ce{CO} consumes all C and \ce{H2O} takes the leftover O atoms. This gives $x_{\ce{H2O}}=1.4\times 10^{-3}$ and $x_{\ce{CO}}=3.2\times 10^{-3}$ at the lower boundary, which is consistent with equilibrium chemistry calculated with \texttt{FastChem} \citep{StockEtal2022}.
In \Tb{abundance} and \Fg{paraboundary} we consider two cases with different elemental abundances: increasing the metallicity to 10 times solar values ($10\times$-[Z]); and increasing C/O to $n_C/n_O=0.9$. 

In the $10\times$-[Z] case, the cloud bottom extends deeper in the atmosphere up to 3 bar. This is because the condensation curves (dashed lines in \Fg{TP}) shifts rightwards when increasing vapor concentrations, intersecting the T-P profile at higher temperatures and pressures. For the same reason, the \ce{Mg2SiO4} cloud does not evaporate until $0.3\ \mathrm{bar}$. 
The formation of the low-hanging \ce{Mg2SiO4} cloud consumes almost all \ce{Mg}, causing low \ce{Mg} concentration above $0.03\ \mathrm{bar}$, (see \Fg{paraboundary}a). Therefore a cloud layer dominated by \ce{SiO2} forms on top of this, with its maximum mixing ratio reaching $10^{-3}$. 
The cloud particles sizes in the $10\times$-[Z] run become over an order of magnitude larger than in the default model (\Fg{default}) with sizes in Fe-dominated region reaching ${\approx}70\ \mathrm{\mu m}$. The large size leads to fast settling, resulting in a thinner iron layer and a razer-thin refractory cloud deeper in the atmosphere (around $2\ \mathrm{bar}$). 

Though all molecular mixing ratios at the lower boundary increase by a factor of 10, we do not see a concomitant growth of the cloud in terms of column density. In fact only 3 times more cloud (in terms of column density) is formed compared to the default case. The reason is that settling provides a negative feedback to cloud growth: the more condensates formed force the cloud to settle faster to hotter, lower region, giving negative feedback to the cloud formation. 
Interestingly, above $2{\times }10^{-2}\ \mathrm{bar}$ the Mg-Si cloud mixing ratio and particles sizes remain unchanged compared to the default solar metallicity run. Although the planet bulk vapor abundances are enhanced, the excess vapor is largely consumed by the "deep" dense cloud and is unable to reach the extended upper cloud ($P<2\times 10^{-2}\ \mathrm{bar}$). 
The decoupling of metallicity and cloud concentration (especially the upper cloud concentration) suggests that even with the presence of clouds, a metal-rich planet may yet exhibit strong molecular lines, which we will demonstrate further in \Se{spectrum}. 

We also ran a model in which the C/O ratio is increased such that $n_{\ce{C}}/n_{\ce{O}}=0.9$, but where the oxygen plus carbon elemental abundance by mass, $x_{\ce{C}}+x_{\ce{O}}$ remains at solar values (\Fg{paraboundary}b). As a result, \ce{CO} consumes most of the oxygen, decreasing the \ce{H2O} abundance at the lower boundary to $x_\mathrm{\ce{H2O}}=6.8\times 10^{-5}$, i.e.,   
less water is present in the atmosphere. 
The scarcity of \ce{H2O} suppresses the formation of Mg-Si, \ce{Al2O3} and \ce{TiO2} cloud particles, which all require water vapor (\Tb{reactions}). As a result, \ce{Fe} becomes the dominant cloud species (\Fg{paraboundary}b). A similar trend has been reported by \citet{HellingEtal2017}
It is of course well-known that C/O=1 constitutes a clear boundary above which the chemistry of the atmosphere and composition of the cloud switches from O-rich to C-rich. But here we find that even before C/O${=}1$ is reached the cloud composition shows a clear dependence with increasing envelope bulk C/O, becoming \ce{Fe}-rich.

As the surface tension of \ce{Fe} \citep[$1870\ \mathrm{dyn}\ \mathrm{cm}^{-1}$, ][]{BrilloEgry2005} is higher than silicates \citep[$307\ \mathrm{dyn}\ \mathrm{cm}^{-1} $ for \ce{SiO2}, ][]{Janz1967}, \ce{Fe} vapor can only condense onto larger nuclei due to the Kelvin effect (\eq{Kelvin}).
Therefore, the nuclei particles need to coagulate before \ce{Fe} can condense onto them. To mimic a more realistic nucleation profile when the particle is made primarily of \ce{Fe}, we tested the C/O=0.9 case where larger nuclei particles are injected into the deeper atmosphere. We take the nuclei particle size $a_{p0}=5\ \mathrm{nm}$, a factor of 5 higher than the default parameter and nucleation pressure $P_n=6\times 10^{-4}\ \mathrm{bar}$, an order of magnitude higher. The result is shown in \Fg{paraboundary}c. 
As explored in \Se{nucleation}, the nucleation profile has minimal effect on the cloud formation, except that the clouds hang lower in the atmosphere. Larger particle size make \ce{Fe} particles harder to diffuse up to the Mg-Si condensing region. Therefore, the upper cloud is composed of comparatively more Mg-Si, though \ce{Fe} is still the dominant species, unaffected by the choice of nucleation height and size.

\subsection{Synthetic Spectrum}\label{sec:spectrum}

\begin{figure*}
    \includegraphics[width=\textwidth]{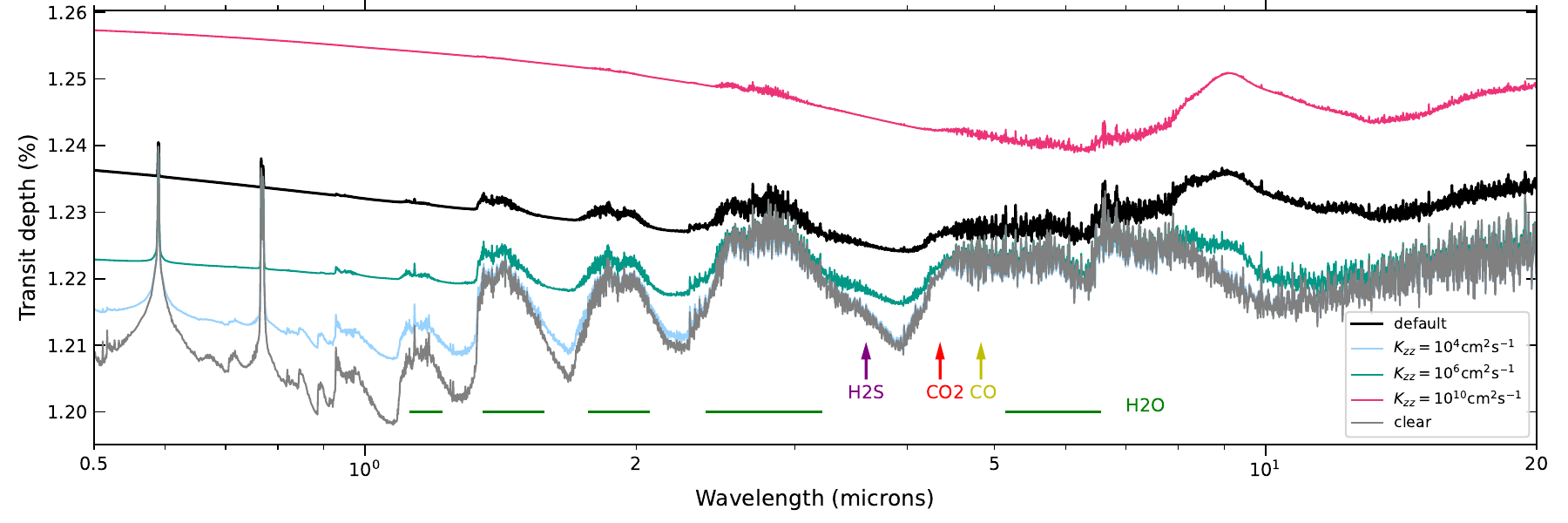}
    \caption{\label{fig:spectrum} 
    Transmission spectrum of a cloudy hot Jupiter (black line). The grey line shows the transmission spectrum of a clear atmosphere, leaving the gas components unchanged. $K_{zz}$ is varied to show how cloud formation changes the atmospheric transmission spectrum. The horizontal lines and arrows indicate molecular features. At long wavelengths the $K_\mathrm{zz}=10^4\,\mathrm{cm^2\,s^{-1}}$ run curves are indistinguishable from the ``clear'' curve. }
\end{figure*}

In \Fg{spectrum} we plot the transmission spectra of the default hot Jupiter model (black curve), focusing on $0.5\ \mathrm{\mu m}$ to $20\ \mathrm{\mu m}$, which covers the wavelength of HST, Spitzer and JWST. 
To demonstrate the contribution of clouds, we also plot in grey the spectrum that would emerge when the solid particles are removed, while preserving the gas species. 
Compared with the clear model, the presence of clouds enhances the transit depth on the order of ${\sim}10^2\ \mathrm{ppm}$. At the short wavelength end, the cloudy spectrum shows a slope consistent with Rayleigh scattering, reflecting the presence of sub-$\mathrm{\mu m}$ size particles.
Furthermore, a distinct silicate feature shows up at 8 to 10 $\mathrm{\mu m}$, unambiguously indicating the formation of silicate particles. 
Though \ce{Fe}, \ce{Al2O3} and \ce{TiO2} cloud layers also form (\Fg{default}), they lie under the $\tau=1$ surface and are thus obscured by the magnesium-silicate cloud layer.

Without clouds the amplitude of the molecular lines can reach $100$ to $200\ \mathrm{ppm}$. The thick clouds in the atmosphere reduce the amplitude to ${\sim}50\ \mathrm{ppm}$. From the spectrum we can identify the \ce{CO} feature at $4.6\ \mathrm{\mu m}$ and the \ce{H2O} features at $2.5$ and $5.6\ \mathrm{\mu m}$, though cloud formation depletes the \ce{H2O} mixing ratio from its boundary value ($1.4\times 10^{-3}$) to $x_{\ce{H2O}}\approx 8\times 10^{-4}$ (\Fg{default}).
Besides the $2.9\ \mathrm{\mu m}$ \ce{H2O} line, \ce{H2S} can still be marginally identified as a shoulder to the right, at $3.5\ \mathrm{\mu m}$.
On the other hand, the features of \ce{CH4} are not found in the spectrum, because most of the carbon exists in the form of \ce{CO} in the chemical equilibrium state for the temperatures considered here.

\begin{figure*}
    \includegraphics[width=\textwidth]{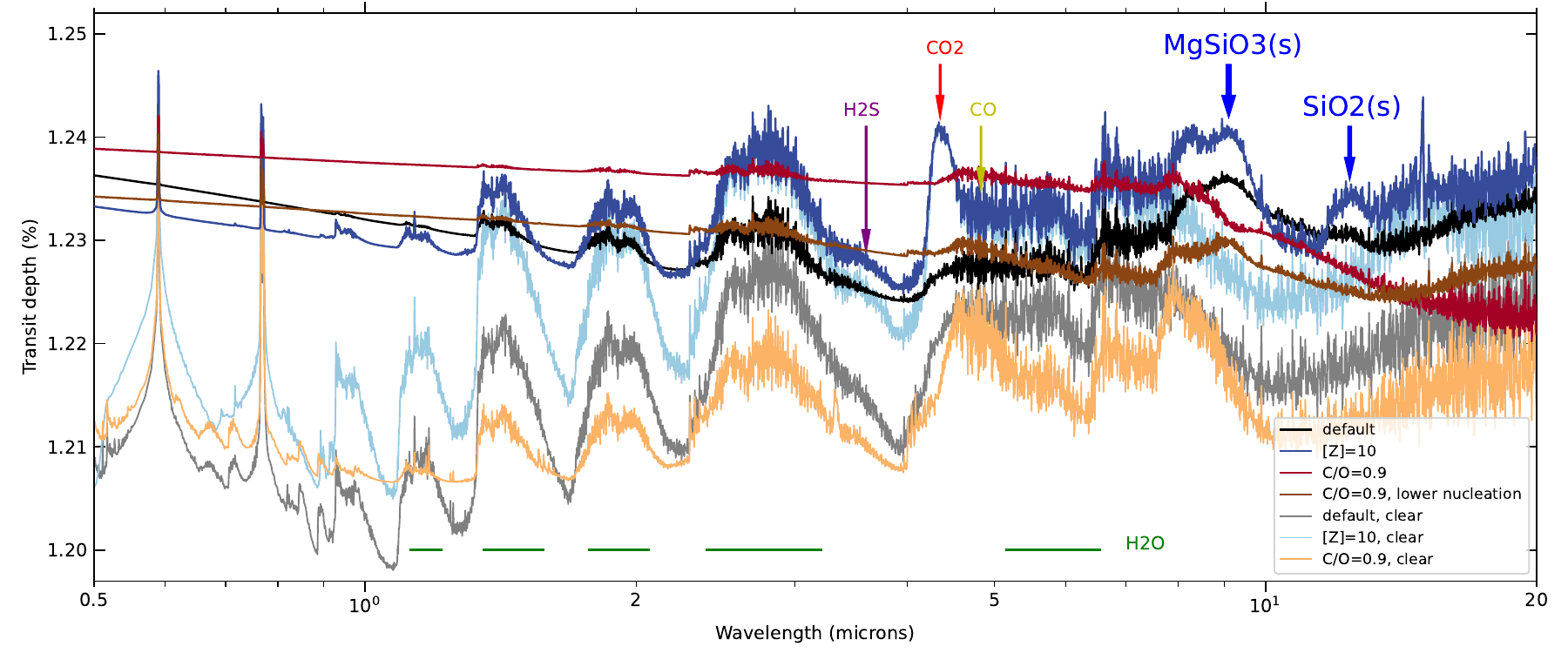}
    \caption{\label{fig:spectrumpara} 
    Transmission spectrum of models varying metallicity and C/O ratio, with dark and light line colors referring to cloudy and clear atmosphere, respectively. The black and grey lines are the default cloudy and clear model (\Se{cloudsolid}). $10\times$-[Z] and C/O=0.9 cases are shown by the blue lines and the red lines. For the C/O=0.9 run, we additionally conduct a run injecting larger ($5\ \mathrm{nm}$) nuclei particles deeper in the atmosphere to accounting for the Kelvin effect (brown curve).
    }
\end{figure*}

We further show in \Fg{spectrum} how $K_{zz}$ affects the spectrum. As we have discussed in \Se{parameterstudy}, the higher the $K_{zz}$, the thicker the cloud, suppressing the molecular features. Indeed we see that the scattering slope dominates the entire spectrum below $5\ \mathrm{\mu m}$ when $K_{zz}=10^{10}\ \mathrm{cm^2\,s^{-1}}$. The $10\ \mathrm{\mu m}$ silicate bump also becomes more prominent.
With decreasing $K_{zz}$, the water lines in the near-IR band become more and more prominent, together with the $3.5\ \mathrm{\mu m}$ \ce{H2S} feature because the cloud becomes thinner. Decreasing $K_\mathrm{zz}$ further to the extremely low $10^4\,\mathrm{cm^2\,s^{-1}}$ even the $10\ \mathrm{\mu m}$ silicate feature vanishes. The silicate cloud lies so deep in the atmosphere that the gas optical depth 
from $2$ to $20\,\mu\mathrm{m}$ above it already exceeds 1. However, clouds still make a difference in the transit depth at wavelengths smaller than $2\ \mathrm{\mu m}$, because the cloud opacity still contributes to the total opacity, raising the height where the total optical depth reaches unity.

We show how the boundary elemental abundance affects the spectrum in \Fg{spectrumpara}.
When the metallicity is increased by a factor of 10, the \ce{H2O}, \ce{H2S} and \ce{CO2} concentrations are boosted, but not much more silicate cloud particles form in the upper atmosphere (\Se{bulk}). As a result, these molecular features become more prominent (see \fg{spectrumpara}). The most prominent example is \ce{H2O} features at $1.4$, $1.9$ and $2.8\ \mathrm{\mu m}$. As the cloud density and particle sizes in the optically thin region (above $10^{-2} \ \mathrm{bar}$) closely resemble the default run (see \Se{bulk}), the $10\ \mathrm{\mu m}$ feature strength remain unchanged. 
Another remarkable result of the $10\times$-[Z] run is the formation of a \ce{SiO2} cloud in the upper atmosphere (\Fg{paraboundary}a, also \citet{WoitkeEtal2020}). As a result, a $12\ \mathrm{\mu m}$ solid feature arises due to \ce{SiO2(s)}. 

For the C/O=0.9 case, there is little \ce{H2O} that can supply the formation of Mg-Si clouds (see \Se{bulk}). Therefore the \ce{MgSiO3} and \ce{SiO2} features around $10\ \mathrm{\mu m}$ disappear (\Fg{spectrumpara}, red curve). 
The opacity increases due to higher \ce{Fe} mixing ratio in the upper atmosphere. Compared to the default run, the reduced amount of Mg-Si solid leads to weaker settling, hence retaining more \ce{Fe} in the upper atmosphere.
In addition, because the \ce{Fe}-to-MgSi ratio is higher than the default case, the effective refractive index of cloud particles is closer to that of \ce{Fe}, showing strong absorption at near-IR wavelengths.
As a result the $\tau=1$ surface is raised to $6\times 10^{-3}\ \mathrm{bar}$ at wavelengths smaller than $5\ \mathrm{\mu m}$ where most molecular features lie. 
However, the opacity in the default model may be somewhat suppressed due to the assumptions inherent to the Bruggeman mixing rule with spherical inclusions (\Se{RT}). 
Featureless near-IR transmission spectra of hot Jupiter planets could be attributed to the presence of \ce{Fe} cloud particles or high turbulent diffusivity $K_{zz}$, but mid-IR spectra at $10\ \mathrm{\mu m}$ wavelength can help distinguish the two scenarios. \
When injecting nuclei deeper in the atmosphere, the near-IR part of the spectrum shows less transit depth, but the shape and amplitude of the spectral features is similar to the C/O=0.9 case (red curve). At $10\ \mathrm{\mu m}$, the Mg-Si features becomes more apparent. The reason for these changes are that the cloud is located deeper in the atmosphere and that its particles are composed of more transparent Mg-Si material.

\section{Application to Self-Luminous Planets and Sub-Neptunes} \label{sec:application}
\codename is designed to be broadly applicable. In this section, we demonstrate the adaptability of \codename by applying it to a self-luminous planet and a sub-Neptune planet, with the aim to show how clouds affect the emission and transmission spectrum.
\subsection{Self-luminous Planet}\label{sec:HR8799}
\begin{figure*}
    \includegraphics[width=\textwidth]{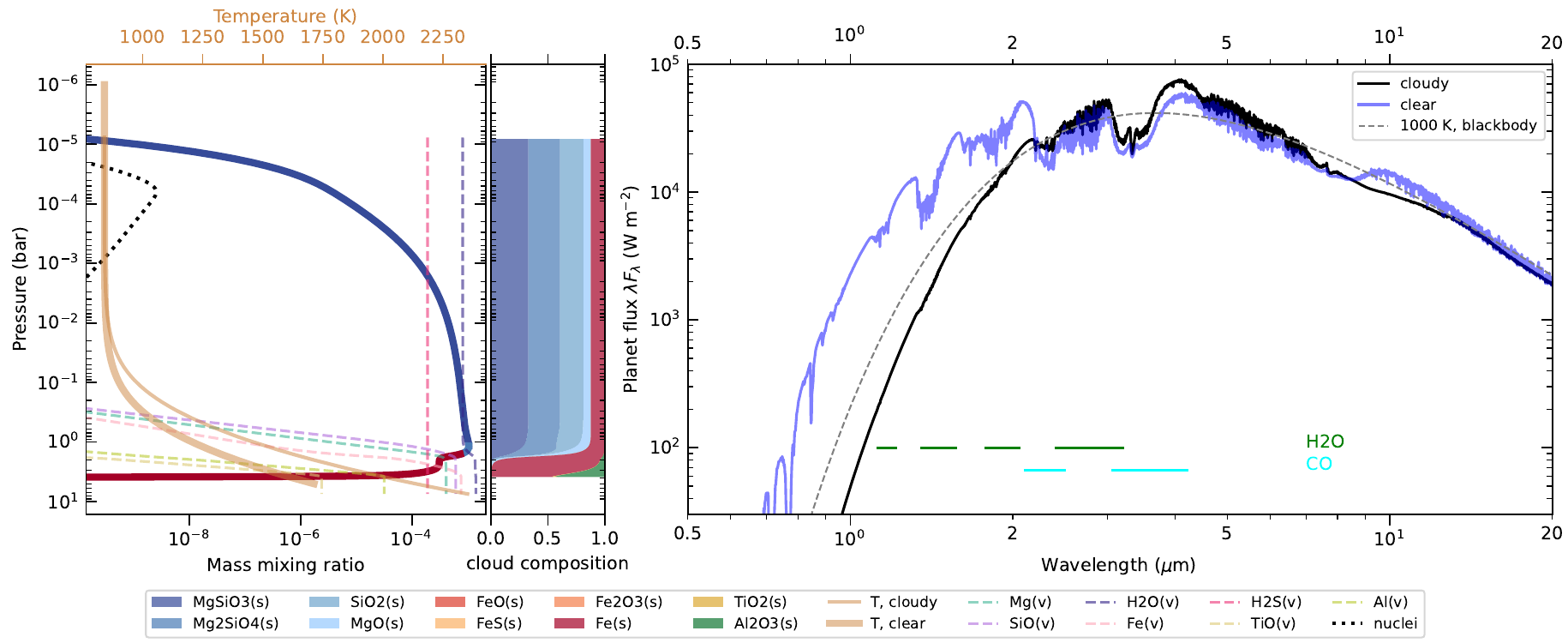}
    \caption{\label{fig:HR8799e}
    Cloud structure and emission spectrum of a self-luminous planet with physical properties similar to HR 8799 e. The left panel shows the cloud, gas and nuclei mixing ratios, with every components the same meaning as \Fg{paraboundary}. The brown thick and thin lines present the temperature profiles we used in the clear and cloudy models, respectively.
    The right panel shows the synthetic emission spectrum of a cloudy atmosphere (the black line) and a clear atmosphere (the blue line), whose effective temperatures are both $1000\ \mathrm{K}$. For reference, the spectrum of a 1000 K black body radiation is also plotted as the dashed line.
    }
\end{figure*}

Self-luminous planets are planets with a temperature structure entirely governed by their internal thermal flux. 
Detected through direct imaging, these planets are relatively massive and orbit far from their host stars ($P\gtrapprox 10\ \mathrm{yr}$). 
These planets are typically young (${\lessapprox}100\ \mathrm{Myr}$) and are still in the process of radiating away their formation energy. Self-luminous planets therefore have a hot interior and much cooler atmosphere. Consequently, we expect clouds to form. When the effective temperature is fixed, the presence of clouds would make the spectrum redder \citep{CurrieEtal2023}. For example, the strong emission of 51 Eri b in Lp ($3.8\ \mathrm{\mu m}$) band, probed by Gemini, lends support to a thick cloud with small particle size \citep{MacintoshEtal2015, SamlandEtal2017}. 

To explore how clouds influence the emission spectrum of self-luminous planets, we apply \codename to a $10\ M_J$ distant giant planet akin to HR 8799 e \citep{GravityCollaborationEtal2019}.  We take the \citet{Guillot2010} temperature profile for a self-luminous planet, where the outgoing irradiation flux determines the temperature profile.The controlling parameters here is the internal irradiation temperature $T_\mathrm{int}$ and the constant, grey opacity at the irradiation IR band $\kappa_\mathrm{IR}$.  For the former we take the internal irradiation flux corresponding to a temperature of $1000\ \mathrm{K}$ \citep{BonnefoyEtal2016, GravityCollaborationEtal2019}.
If a self-luminous planet atmosphere, cloudy or not, is in thermal equilibrium, it will radiate away the same amount of energy it receives from the interior.
To isolate the role clouds play in a self-luminous planet, we tuned the IR opacity parameter in the Guillot model, such that the effective temperature $T_\mathrm{eff}$ at which the planet (cloudy or clear) radiates is the same as $T_\mathrm{int}=1000\ \mathrm{K}$.  For a clear atmosphere, $T_\mathrm{int}=T_\mathrm{eff}$ translates to $\kappa_{IR}=0.04\ \mathrm{cm}^2\ \mathrm{g}^{-1}$ and for a cloudy atmosphere this gives $\kappa_{IR}=0.1 \mathrm{cm}^2\ \mathrm{g}^{-1}$, consistent with the addition of the cloud opacity.
The temperature profile of the cloudy run is shown in \Fg{TP}. As the lower atmosphere can be as hot as $2000\ \mathrm{K}$, the condensate species and reactions are the same as listed in \Tb{reactions}. Here, the goal is not to \textit{fit} the observed spectrum of HR 8799 e in particular, but rather to demonstrate how \codename can be applied towards self-luminous planets and how clouds change the planet's emission spectrum in general.

The cloud structure and emission spectrum calculated from \texttt{petitRADTRANS} are shown in \Fg{HR8799e}. Similar to the hot Jupiter, the self-luminous planet clouds also show a layered structure. The upper cloud above $1.5\ \mathrm{bar}$ is dominated by magnesium silicate. Due to the lack of stellar irradiation heating, most of the region where clouds form remains cooler than $1000\ \mathrm{K}$. Therefore magnesium-silicate particles do not evaporate until $1.5\ \mathrm{bar}$ ($1600\ \mathrm{K}$). 
With increasing pressure, the temperature quickly increases with depth, reaching the iron evaporation temperature at $3.5\ \mathrm{bar}$. Consequently, iron becomes the dominant cloud species in the $1.5\ \mathrm{bar}$ to $3.5\ \mathrm{bar}$ range. But we also note that due to diffusion, \ce{Fe} is efficiently transported to the magnesium-silicate cloud region, comprising 15\% of the cloud particle mass there.

In the right panel we show the synthetic emission spectrum. The spectrum of a clear atmosphere devoid of clouds, and a cloudy atmosphere are drawn with the blue and black lines, respectively. 
The emission spectrum of the clear self-luminous planet resembles a black-body with additional molecular absorption features. The absorption lines arise because the photons from the hotter interior get absorbed by molecules in the cooler upper region.

In the cloudy run, photons originating from the hot interior are absorbed by the cloud particles in the atmosphere and re-emitted at a cooler emission temperature. Therefore, the  spectrum is redder compared to the clear one, even though the effective temperature of the two atmospheres are the same.
Moreover, the molecular lines ${<}2\ \mathrm{\mu m}$ are suppressed due to cloud absorption. We observe absorption lines of \ce{CO} and \ce{H2O} only in the near-IR band. For example, the dip from 3 to 4 $\mathrm{\mu m}$ is due to absorption of \ce{CO}.
The color of the spectrum and the line strength at near-IR wavelengths allow us to distinguish a cloud-free self-luminous planet from a cloudy one. 
Though fitting a real spectrum is beyond the scope of this work, we note that SPHERE \citep{ZurloEtal2016}, GPI \citep{GreenbaumEtal2018} and VLTI/GRAVITY \citep{GravityCollaborationEtal2019, MolliereEtal2020, NasedkinEtal2024} spectrum of HR8799 e lies in between our cloudy and clear cases, suggesting less thick cloud as our cloudy run, e.g. smaller $K_{zz}$.

\subsection{Sub-Neptunes}\label{sec:GJ1214}
\begin{figure*}
    \includegraphics[width=\textwidth]{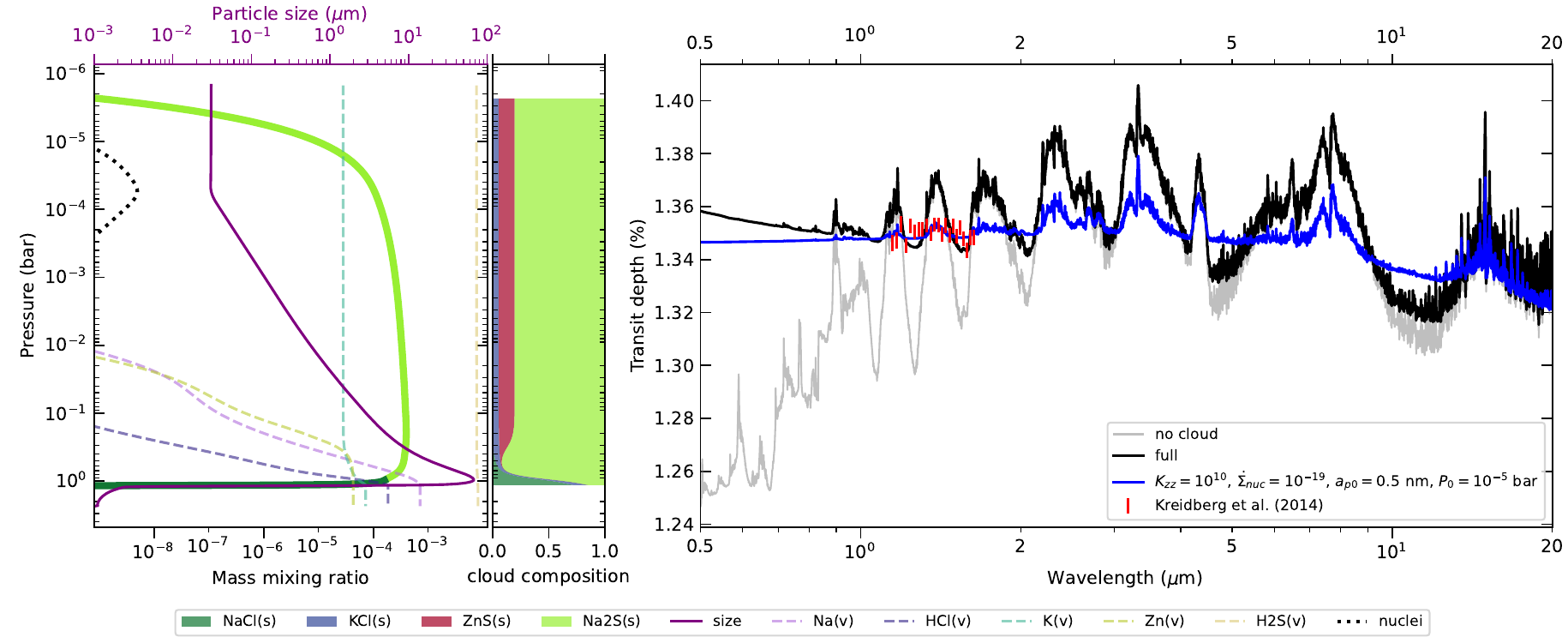}
    \caption{\label{fig:subNeptune}
    The cloud structure and transmission spectrum of a sub-Neptune planet with physical properties similar to GJ 1214 b. The condensate species considered here is \ce{NaCl}, \ce{KCl} and \ce{ZnS}. In the left panel, the meaning of each line is the same as \Fg{paraboundary}.
    The cloud is dominated by \ce{Na2S} (upper part) and \ce{NaCl} (cloud bottom). The right panel shows the synthetic spectrum and the blue line shows one set of parameters that leads to flat spectrum in $1$ to $2\ \mathrm{\mu m}$ range, consistent with observation \citep{KreidbergEtal2014}.}
\end{figure*}

In this section we focus on a warm sub-Neptune planet, whose temperature and mass is lower than that of a hot Jupiter. As the condensation temperatures of magnesium-silicates are always far higher than the local atmospheric pressure (see \Fg{TP}), the condensates in \Tb{reactions} are likely to form deeper in the interior or near the surface of the planet.
The condensates formed in the atmosphere regions of sub-Neptune planets are usually salts, like \ce{KCl} and \ce{NaCl} \citep{MorleyEtal2013, OhnoOkuzumi2018}. 

Here we consider a sub-Neptune with physical properties similar to GJ 1214 b, a $8.17\ M_\oplus$ and $2.74\ R_\oplus$ planet $0.015\ \mathrm{au}$ away from a $0.216\ R_\odot$ star \citep{CarterEtal2011, CloutierEtal2021, MahajanEtal2024}, with irradiation temperature at $300\ \mathrm{K}$. 
JWST observations of GJ 1214 b reveal thick and reflective cloud layers \citep{KemptonEtal2023}.
Thus the condensate species chosen are \ce{NaCl}, \ce{KCl}, \ce{Na2S} and \ce{ZnS}, whose condensation temperatures lie in the atmospheric temperature range \citep{MorleyEtal2013, CharnayEtal2015, GaoBenneke2018}. We take a 100 times solar metallicity (\Tb{abundance}), with \ce{Na}, \ce{K}, \ce{Cl}, \ce{Zn} and \ce{S} existing in the form of \ce{Na}, \ce{K}, \ce{HCl}, \ce{Zn}, \ce{H2S} vapor. 
The chemical reactions that form these species are listed in the lower part of \Tb{reactions}.
Other parameters, like $K_{zz}$ and the nucleation description, are kept the same as \Tb{parameters}, except the mean molecular weight $m_\mathrm{gas}$, for which 5m$_\mathrm{u}$ is taken, consistent with its metallicity.

The atmosphere structure and the resulting transmission spectrum for the sub-Neptune planet are shown in \Fg{subNeptune}. As a result of the solar element ratios, \ce{HCl} is the limiting species for \ce{NaCl} and \ce{KCl}, and \ce{Na} is limiting the formation of \ce{Na2S}. We find that \ce{Na}-salts are the major cloud species, because \ce{Na} is over one order of magnitude more abundant than \ce{K} or \ce{Zn}. 
In the right panel of \Fg{subNeptune}, we show the transmission spectrum from visual to mid-IR band. Although scattering masks the molecular features for wavelengths $\lambda<1\ \mathrm{\mu m}$, various features from \ce{CO2} and \ce{CH4} unveil themselves at IR wavelengths. 
Unlike the magnesium-silicates on hot Jupiters, \ce{Na}-salts are mostly transparent at $\lambda>2\ \mathrm{\mu m}$ and thus the spectrum at longer wavelength is insensitive to the formation of the \ce{Na2S} and \ce{NaCl} salts.

The Hubble Space Telescope found that the warm-Neptune GJ 1214 b shows a flat spectrum from 1 to 2 $\mathrm{\mu}$m, with an amplitude of less than 100 ppm \citep{KreidbergEtal2014}. The HST observation is plotted in \Fg{subNeptune}. We further tested how \codename can reproduce a similarly flatness of the spectrum, though our intention is not to quantitatively retrieve the planet property in this paper. With higher $K_{zz}$ and smaller $a_{p0}$, we find that thicker clouds form and extend to the upper atmosphere.
In addition, when decreasing the nucleation rate by four orders of magnitude, the particle sizes becomes larger, reaching micron size throughout the cloud. We also changed the reference pressure of the transmission spectrum calculation to $10^{-5}\ \mathrm{bar}$ The combined effect of decreasing $\dot{\Sigma}_\mathrm{nuc}$ and increasing $K_\mathrm{zz}$ is that clouds with larger particle sizes becomes dominant in the upper atmosphere, which is consistent with a flat spectrum with low amplitude \citep{KemptonEtal2023}.

\section{Discussion} \label{sec:discussion}
\subsection{Assessment on the performance and assumptions}
In the preceding sections we have demonstrated the viability of \codename to calculate the cloud distribution and to obtain realistic synthetic model spectra under different planet conditions. For \codename the simplifications pertain primarily the prescription of nucleation at the expense of a number of free parameters ($a_{p0}$, $\dot{\Sigma}_n$ and $\sigma_n$), in addition to the more standard assumptions of 1D T-P profile and fixed $K_{zz}$. 
Under these assumptions, \codename computes the mixing ratio of cloud and vapor, according to realistic condensation prescription in tandem with particle transport. Indeed, \codename is efficient. On average, each complete iteration for the default hot Jupiter (\Tb{reactions}), with 100 grid points, takes 2.1 seconds on a laptop PC (Intel Core™ i7-12700H @4.7GHz). 
In the following two paragraphs, we will assess the assumptions adopted by \codename, aiming to answer the question: how reliable is \codename's result.

One caveat of this work is the assumption of a single representative size for the particles at each height. Though this assumption works for hot Jupiters where nuclei form in the higher atmosphere due to, e.g. nucleation of refractory species like \ce{TiO2} \citep{LeeEtal2015, HellingEtal2017, Helling2019} or photochemistry \citep{ZahnleEtal2016, KawashimaEtal2019}, it may not be compatible with the nucleation condition on a terrestrial planet characterized by a surface. 
On Earth, nuclei are supplied from the surface by the evaporation of ocean salt particles or the eruption of volcanoes \citep{Woods1993, QuinnEtal2017}. These nuclei diffuse up due to turbulent motion, until growing too large due to vapor condensation and precipitate down to the surface. 
Therefore, the ground level is characterized by two particle sizes: one for smaller nuclei, and the other for the larger rain drops. At present, such a bimodal size distribution cannot be accounted for in the current version of \codename. One possible solution is to model the size of the particles with two or more bins. For example, \citet{OhnoOkuzumi2017} followed the smaller ``cloud particles'' and larger ``rain particles'' separately, which by their definition, have upward and downward net velocity, respectively. 
The smaller particles are carried up by the background upwelling vapor, until they grow heavy enough by condensation to rain down, at which point they are "promoted" to rain particles. These large particles will sweep-up the smaller particles during their descent. 
In studies that model the size distribution of particles in protoplanetary disks, such a two-population size distribution has also been widely applied and has proven to be in good agreement with the full size-distribution simulations \citep{BirnstielEtal2012, BirnstielEtal2015}. In future works, \codename can be adjusted to a similar strategy, solving the transport and formation equation of both small nuclei and large particles. 

Another key assumption of \codename is the single spatial dimension. In reality, horizontal variation of temperature and transport are important factors that shape the appearance and properties of clouds. 
For example, hot Jupiters are usually tidally locked, with a hot day-side and a cooler night-side.  It is found that due to lower irradiation intensity, clouds are prone to form in the higher latitude region or the night side \citep{LinesEtal2018, RomanRauscher2019, RomanEtal2021}
As a result, caution should be exercised when modeling tidally-locked planet spectra or phase curves with a horizontally isothermal temperature model. Alternative to modelling clouds in 2D or 3D, which takes significantly more computational resources, one can set up a  collection of 1D vertical grids spanning the equatorial plane or the entire surface \citep{WebberEtal2015,  FengEtal2020, HellingEtal2020, SamraEtal2023}. With independent temperature profiles for each vertical model, \codename can generate a 2D or 3D cloud profile resulting in a more realistic spectrum.
In a similar vain, the horizontal transport of cloud particles and vapor could be more efficient than vertical diffusion, as is suggested by self-consistent 2D simulation \citep{PowellZhang2024}. Therefore, the horizontal transport of vapor replenishes the cloud-depleted region, increasing the cloud formation efficiency. The horizontal advection can also transport cloud particles to hot regions where 1D models would predict clouds to be absent. 
Similar to \citet{PowellZhang2024}'s approach, a collection of \codename clouds on the equatorial plane can be post-processed with horizontal advection, under a velocity characterizing the longitudinal jet. Though not fully modelling the 3D nature, this approach serves as a reasonable first-order approximation that empowers the atmosphere retrieval with horizontal transport of the cloud. 

\subsection{The need for a self-consistent clouds in retrievals}
Several works have modelled cloud formation with a higher degree of physical consistency. For example, \texttt{DRIFT-PHOENIX} \citep{WitteEtal2009} considers nucleation rates consistent with the the vapor abundance and calculates gas phase equilibrium chemistry on the fly. Following this approach, \texttt{DIFFUDRIFT} \citep{WoitkeEtal2020} accounts for vertical transport of dust and gas similar to \codename. Besides, \texttt{CARMA} \citep{TurcoEtal1979, ToonEtal1979} solves the particle size distribution with ${\sim}10$ mass bins. 
Though designed for higher precision, the qualitative structure of clouds in hot Jupiters obtained by these codes is very similar to the computations by \codename: Mg-Si at the top, Fe below it and \ce{Al2O3} in the lower regions of the cloud. Despite the similar results, however, the above codes are usually too time-consuming to be directly used in atmosphere retrieval. One \texttt{DRIFT-PHOENIX} run can take 2 CPU minutes and one \texttt{DIFFUDRIFT} run can take 500 CPU hours \citep{WoitkeEtal2020}. Therefore, these models are typically used as pre-calculated forward model grids \citep[e.g.,][]{SamraEtal2020, GaidosHirano2023, YangEtal2024}. In contrast, \codename has been designed to trade CPU-consuming precision for efficiency, rendering it suitable for retrievals.

With \codename, we are able to retrieve the transmission spectrum in a self-consistent way. As clouds can suppress molecular features and change the molecular abundances in the upper atmosphere, cloud-free retrieval results may not represent the bulk abundances of the planet. Specifically, the molecular abundances may be underestimated, or the mean molecular weights may be overestimated. 
In addition, retrievals where parameterized clouds are put in by hands introduce excessive parameters, which cause degeneracies among the parameters \citep[e.g. ][]{BurninghamEtal2017}, resulting in poorly constrained molecular abundances. A joint retrieval, where cloud formation is modeled simultaneously with the gas components in a physical model can lift the degeneracy. 
With an ever-increasing number of multi-band spectra on planet atmospheres, \codename allows us to retrieve the cloud and gas simultaneously, providing a physically consistent posterior on metallicity and C/O ratio. 

The multi species feature of \codename also opens the opportunity to constrain the composition of cloud particles. This is because solid particles of different composition have different optical properties at mid-IR wavelength \citep{GrantEtal2023, DyrekEtal2024}. For example, \ce{SiO2} displays an additional absorption feature at $12\ \mathrm{\mu m}$ compared to \ce{MgSiO3} (see \Se{bulk} and \Fg{spectrumpara}).
Recently, JWST / MIRI observations on WASP-17 b directly observed silicate features at around $10\ \mathrm{\mu m}$ wavelength. The silicate peak in the spectrum clearly collocates with \ce{SiO2} in quartz form \citep{GrantEtal2023}, although a mixture with other silicates is not ruled out. A physically-consistent cloud model is necessary to find out which cloud species form under the bulk condition of these hot planets and offers the opportunity to find a unique solution. 

\subsection{Future directions for model improvement}
There are two directions where \codename can be extended towards greater consistency. First, this paper calculates the T-P profile from the \citet{Guillot2010} model which assumes a two-stream approach with fixed inward (visible) and outward (IR) opacities. 
However, cloud formation alters the optical properties and hence the temperature structure in the atmosphere \citep{RomanRauscher2019, HaradaEtal2021, RomanEtal2021}. Accounting for the cloud opacity during radiative transfer can influence the atmosphere temperature. For example, on the one hand, \citet{RomanRauscher2019} found that scattering by cloud particles has an effect of cooling down the atmosphere. On the other hand, the absorption by cloud particles generally warms up the atmosphere \citep{PhillipsEtal2020, RomanRauscher2019}.

Second, \codename at present does not account for the energetics involved in phase-change processes. On colder planets, cloud formation releases latent heat and would change the temperature profile \citep{JohansenEtal2021, GeEtal2023}. 
The fast computation of a cloud profile by \codename paves the way for a retrieval coupling the T-P profile with clouds \citep{MinEtal2020}. In a future work, iteration between cloud formation and thermal structure calculation through radiative transfer will be performed to comprehensively understand how clouds change the planet spectrum.

\section{Conclusions} \label{sec:conclusion}

We have developed \texttt{ExoLyn}: an exoplanet cloud model, specifically tailored for atmospheric spectrum retrieval. \codename solves the formation and transport equations of multi-species cloud condensates, vapor and particle number (nuclei), under turbulent diffusion and sedimentation. For condensation, we considered actual chemical reactions that lead to the formation of cloud particles and the depletion of vapor. With an averaged computational time of ${\lesssim}2\ \mathrm{CPU\ s}$ for a 100 grid points, \codename is ideal for MCMC simulations in atmospheric retrieval. 
We apply \codename to hot Jupiters, self-luminous planets and sub-Neptunes. Our conclusions are as follows:

\begin{enumerate}
    \item On a hot Jupiter planet three compositionally distinct cloud layers form from top down. These are: the magnesium-silicate cloud layer, the iron cloud layer and the refractory cloud layer. The magnesium-silicate layer is mainly composed of a mixture of \ce{MgSiO3}, \ce{SiO2} and \ce{Mg2SiO4}. The refractory layer is dominated by \ce{Al2O3}. The layered structure is consistent with the condensation sequence of the condensate species. 
    \item The cloud thickness and extent of the cloud are chiefly controlled by the diffusivity ($K_{zz}$ parameter) of the atmosphere. The stronger the diffusion, the thicker the clouds, the larger cloud particles grow, and the stronger cloud layers mix and overlap. The nucleation properties -- which we prescribe in the model -- has weak effects on the cloud profile. In particular, the effect of increasing the nucleation rate is counteracted by particle coagulation.
    \item For hot-Jupiters at diffusivity levels of $K_{zz}=10^8\ \mathrm{cm}^2\mathrm{s}^{-1}$, iron and refractory clouds are not visible because they are located in the optically thick region in the atmosphere. When thicker clouds form at higher diffusivity, the molecular lines (\ce{H2O} and \ce{CO}) are suppressed, but the magnesium-silicate features at 8 to 10 $\mu\mathrm{m}$ become more prominent. On the other hand, when $K_{zz}\leqslant10^{4}\ \mathrm{cm^2\ s^{-1}}$, clouds disappear from the IR transmission spectra (\Fg{spectrum}), because thin clouds form in the molecular optically-thick layer. 
    \item Increasing metallicity leads to significant changes in the composition of cloud particles and vapor components. The formation of \ce{Mg2SiO4} at a depth of $1\ \mathrm{bar}$ depletes the \ce{Mg} vapor concentration in the upper atmosphere. As a result, the dominant cloud species becomes \ce{SiO2} rather than \ce{MgSiO3}. This transition is reflected in an additional \ce{SiO2} feature at $12\ \mathrm{\mu m}$. Due to higher molecular abundances, metal-rich planets exhibit stronger molecular lines, even with the presence of cloud. 
    \item For a hot Jupiter planet with C/O ratio close to unity, iron clouds dominate because the low \ce{H2O} abundance suppresses the formation of O-bearing species. The iron cloud increases the transit depth, suppresses the molecular lines, and the Mg-Si features disappears. 
    \item \codename is also applicable to self-luminous planets and sub-Neptunes. Clouds that form on self-luminous planets redden the spectra and suppress the molecular lines. On sub-Neptunes \ce{Na2S} and \ce{NaCl} clouds form, whose scattering obscures the molecular features below $2\ \mathrm{\mu m}$.  
    ExoLyn predicts a flat spectrum in the near-IR when a modest amount of nuclei form in a turbulent atmosphere (high $K_\mathrm{zz}$),  consistent with the observation of the sub-Neptune GJ 1214 b.
\end{enumerate}

\section*{Acknowledgements}
This work is supported by the National Science Foundation of China under grant No. 12250610189 and 12233004. The authors thank the anonymous referee for their comments, which have improved the quality of the manuscript. The authors gratefully thank Kazumasa Ohno, Peter Gao, Michael Zhang, Xi Zhang and Paul Molli\`{e}re for insightful discussion.

\section*{Data Availability}
The data underlying this article will be shared on reasonable requests to the corresponding author.
The code \codename is available at \url{https://github.com/helonghuangastro/exolyn}.



\bibliographystyle{aa}
\bibliography{ads} 





\end{document}